\documentclass[aps,prx,article,twocolumn,superscriptaddress]{revtex4-2}
\usepackage{graphicx,wrapfig}
\usepackage{fancyhdr}
\usepackage[colorlinks,citecolor = blue,linkcolor=blue]{hyperref}
\usepackage{ifthen}
\usepackage{color}
\usepackage{epsfig}
\usepackage{dcolumn}
\usepackage{float}
\usepackage{mathtools}
\usepackage{hyperref}
\usepackage{comment}
\usepackage{bbm}
\usepackage{amsmath}

\newcommand{\TNS}{{Ta$_2$NiSe$_5$}}

\begin{document}

\title{Role of electron-phonon coupling in excitonic insulator candidate Ta$_2$NiSe$_5$}

\author{Cheng Chen}
\email{These authors contributed equally to this work}
\affiliation{Department of Physics, University of Oxford, Oxford, OX1 3PU, United Kingdom}
\affiliation{Department of Applied Physics, Yale University, New Haven, Connecticut 06511, USA}

\author{Xiang Chen} 
\email{These authors contributed equally to this work}
\affiliation{Physics Department, University of California, Berkeley, California 94720, USA}
\affiliation{Materials Science Division, Lawrence Berkeley National Lab, Berkeley, California 94720, USA}

\author{Weichen Tang} 
\affiliation{Physics Department, University of California, Berkeley, California 94720, USA}
\affiliation{Materials Science Division, Lawrence Berkeley National Lab, Berkeley, California 94720, USA}

\author{Zhenglu Li} 
\affiliation{Physics Department, University of California, Berkeley, California 94720, USA}
\affiliation{Materials Science Division, Lawrence Berkeley National Lab, Berkeley, California 94720, USA}

\author{Siqi Wang} 
\affiliation{Department of Applied Physics, Yale University, New Haven, Connecticut 06511, USA}

\author{Shuhan Ding} 
\affiliation{Department of Physics and Astronomy, Clemson University, Clemson, South Carolina 29631, USA}

\author{Zhibo Kang} 
\affiliation{Department of Applied Physics, Yale University, New Haven, Connecticut 06511, USA}

\author{Chris Jozwiak}
\affiliation{Advanced Light Source, Lawrence Berkeley National Laboratory, Berkeley, California 94720, USA}

\author{Aaron Bostwick}
\affiliation{Advanced Light Source, Lawrence Berkeley National Laboratory, Berkeley, California 94720, USA}

\author{Eli Rotenberg}
\affiliation{Advanced Light Source, Lawrence Berkeley National Laboratory, Berkeley, California 94720, USA}

\author{Makoto Hashimoto}
\affiliation{Stanford Synchrotron Radiation Lightsource, SLAC National Accelerator Laboratory, Menlo Park, CA 94025, USA}

\author{Donghui Lu}
\affiliation{Stanford Synchrotron Radiation Lightsource, SLAC National Accelerator Laboratory, Menlo Park, CA 94025, USA}

\author{Jacob P.C. Ruff}
\affiliation{Cornell High Energy Synchrotron Source, Cornell University, Ithaca, New York 14853, USA}

\author{Steven G. Louie} 
\affiliation{Physics Department, University of California, Berkeley, California 94720, USA}
\affiliation{Materials Science Division, Lawrence Berkeley National Lab, Berkeley, California 94720, USA}

\author{Robert J. Birgeneau}
\affiliation{Physics Department, University of California, Berkeley, California 94720, USA}
\affiliation{Materials Science Division, Lawrence Berkeley National Lab, Berkeley, California 94720, USA}
\affiliation{Department of Materials Science and Engineering, University of California, Berkeley, California 94720, USA}

\author{Yulin Chen}
\affiliation{Department of Physics, University of Oxford, Oxford, OX1 3PU, United Kingdom}

\author{Yao Wang} 
\email{yaowang@g.clemson.edu}
\affiliation{Department of Physics and Astronomy, Clemson University, Clemson, South Carolina 29631, USA}

\author{Yu He}
\email{yu.he@yale.edu}
\affiliation{Department of Applied Physics, Yale University, New Haven, Connecticut 06511, USA}

\date{\today}

\begin{abstract}
Electron-hole bound pairs, or excitons, are common excitations in semiconductors. They can spontaneously form and ``condense'' into a new insulating ground state -- the so-called excitonic insulator -- when the energy of electron-hole Coulomb attraction exceeds the band gap. In the presence of electron-phonon coupling, a periodic lattice distortion often concomitantly occurs with this exciton condensation. However, similar structural transition can also be induced by electron-phonon coupling itself, therefore hindering the clean identification of bulk excitonic insulators based on reductionistic reasoning (e.g. which instability is the ``driving force'' of the phase transition). Using high-resolution synchrotron x-ray diffraction and angle-resolved photoemission spectroscopy techniques, we identify key electron-phonon coupling effects in a leading excitonic insulator candidate \TNS. These include an extensive unidirectional lattice fluctuation and an electronic pseudogap in the normal state, as well as a negative electronic compressibility in the charge-doped broken-symmetry state. In combination with first principles and model calculations, we determine a minimal lattice model and the corresponding interaction parameters that capture the experimental observations. More importantly, we show how the Coulomb and electron-phonon coupling effects can be separated on the level of lattice model, and demonstrate a general framework beyond the reductionist approach in the investigation of correlated systems with intertwined orders.
\end{abstract}

\maketitle

\section{INTRODUCTION}

Correlated electron systems often contain multiple instabilities and intertwined orders, where numerous degrees of freedom jointly determine the material properties~\cite{fradkin2015colloquium}. A notable example is the iron-based superconductors, where the electronic, orbital and spin channels all contribute to the rotational symmetry-breaking ground state~\cite{fernandes2014drives}. In these systems, order parameters of the same symmetry often couple to each other, and the common reductionistic approach of ascribing the transition to a single dominant channel becomes ambivalent~\cite{schmalian2013settling}. Another more recent example is the renewed search for bulk excitonic insulators, where electron-hole pairs spontaneously form in the ground state and cause charge insulation~\cite{mott1961transition,knox1983introduction}. In indirect band gap semiconductors such as 1T-TiSe$_2$~\cite{kogar2017signatures}, both the plasmon and phonons soften to zero energy below the phase transition temperature, indicating contributions from both the lattice and electronic degrees of freedom. Indeed, it has been extensively discussed that the presence of electron-phonon coupling, regardless of whether there is pre-existing electron-hole Coulomb attraction, can also lead to a structural distortion with an ordering wave vector matching what connects the conduction band bottom and the valence band top~\cite{halperin1968possible,seitz1968solid}. Such a ``chicken-or-egg'' dilemma originates from the pursuit of a reductive description, which is not always possible in correlated materials where relevant interaction energy scales are not well-separated. The lack of a clear energy hierarchy severely limits the construction and application of reductive frameworks. One common manifestation of this issue is the often ambivalent interpretations of observations made via different experimental techniques on the same system. A promising approach to avoid this dilemma is to include all relevant interactions on equal footing in the theoretical model, determine the interaction parameters by comparing the simulation with experimental observables through various tuning methods, and subsequently make further predictions for model validation and for material design guidance~\cite{alexandradinata2020future}.

\begin{figure*}[!t]
\centering
\includegraphics[width= 17 cm]{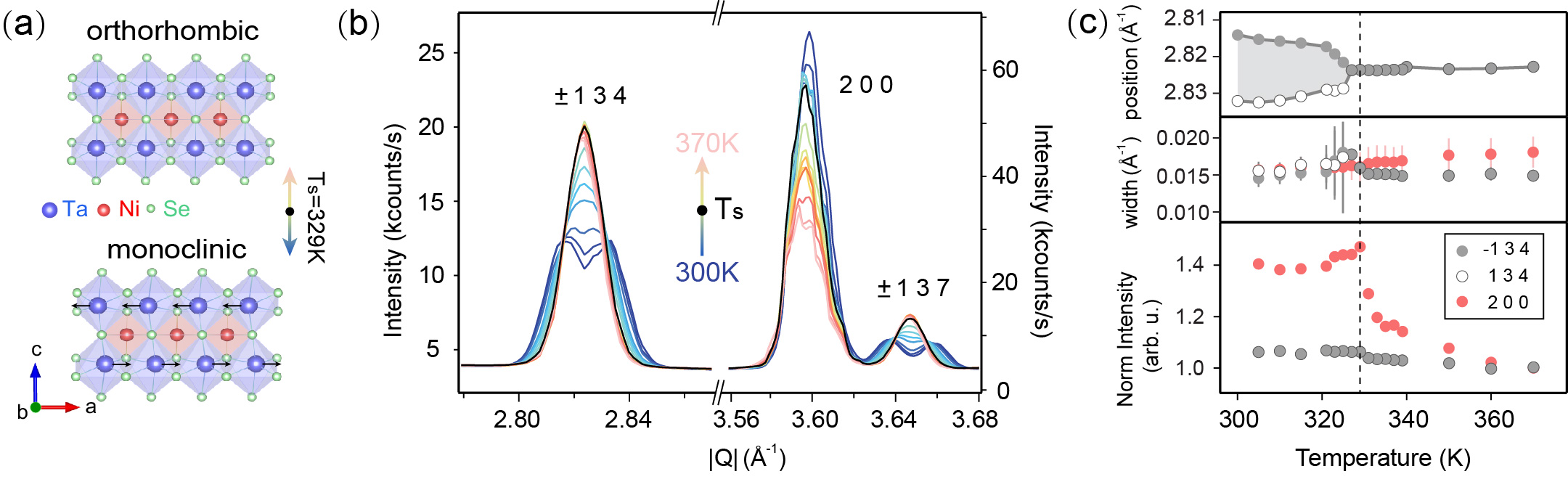}\vspace{-3mm}
\caption{Structural phase transition in \TNS. (a) Top view of the lattice structures across the phase transition: orthorhombic (above $T_s$) and monoclinic (below $T_s$). (b) Temperature dependence of the $\mathrm{\pm}$ 1 3 4, 2 0 0 and $\mathrm{\pm}$ 1 3 7 Bragg peaks from the single crystal x-ray diffraction experiments. (c) Peak positions, widths, and normalized intensities of the $\mathrm{\pm}$ 1 3 4 and 2 0 0 Bragg peaks across the phase transition at $T_s\sim$\,329 K. 
}
\label{fig:Fig1}
\end{figure*} 

In this work, we will adopt such a framework to address the ongoing search for bulk excitonic insulators in low-dimensional materials. Recent studies revealed evidence of exciton formation and strong electron-phonon coupling (EPC) in the quasi-one-dimensional (quasi-1D) ternary chalcogenide \TNS\,\cite{di1986physical, wakisaka2009excitonic,lu2017zero,yan2019strong,volkov2021critical,kim2021direct}. Upon warming, a $q = 0$ monoclinic-to-orthorhombic structural transition happens at $T_s\sim$\,329 K [see Fig.~\ref{fig:Fig1}(a)], above which a semi-metallic electronic structure is supposedly enforced by the mirror symmetry of the crystal and opposite parities of the low-energy orbitals near the Fermi momentum\,\cite{mazza2020nature}. However, an insulating behavior is observed to persist up to 550K\,\cite{di1986physical}. This is in striking contrast to archetypal metal-to-insulator transition (MIT) systems, such as the perovskite nickelates\,\cite{medarde1997structural} and the chain compound TTF-TCNQ\,\cite{chu1973pressure}, where the higher-symmetry structure occurs concurrently with the metallic electronic state (usually called the normal state). The nature of the high-temperature electronic state in \TNS\ remains controversial\,\cite{wakisaka2012photoemission,seki2014excitonic,mor2017ultrafast,okazaki2018photo,tang2020non,baldini2020spontaneous,watson2020band,fukutani2021detecting}, including suggestions of either a regular gapless semimetal or a pseudogapped state with ``preformed excitons''\,\cite{watson2020band,seki2014excitonic,fukutani2021detecting}. In addition, contentions remain regarding whether the system's insulating ground state hosts an exciton condensate\,\cite{wakisaka2009excitonic,lu2017zero,wakisaka2012photoemission,seki2014excitonic,mor2017ultrafast,okazaki2018photo,tang2020non,baldini2020spontaneous,watson2020band,fukutani2021detecting,lee2019strong,larkin2017giant}.

Combining high-energy x-ray diffraction (XRD) and high-resolution angle-resolved photoemission spectroscopy (ARPES) techniques, we first acquire high-resolution experimental results that will be used to test minimal many-body models. The experimental observations mainly focus on normal state properties, where the symmetry is not yet broken. These include an electronic pseudogap state over a wide temperature range above $T_s$ in \TNS, which occurs concomitantly with a strong unidirectional fluctuation of the lattice; a similarly pseudogapped electronic structure during the insulator-to-semimetal transition upon electron injection below $T_s$; and concomitant negative electronic compressibility (NEC), which suggests the thermodynamic inevitability of EPC's contribution. We then combine experimental results with both first-principles and many-body simulations to estimate the microscopic interaction parameters, including the kinetic energy of the conduction and valence electrons, the strength of the Coulomb interaction, and the EPC vertex. Considering both interactions on an equal footing, we discuss the roles of Coulomb interaction and EPC by comparing the experimental data to the numerically simulated spectra with both, either, and none of the interaction terms. Informed by such a comparison, we attribute the pseudogap state and NEC in \TNS\ primarily to the interband electron-hole excitation facilitated by a fluctuating lattice. On the contrary, direct Coulomb attraction between electrons and holes alone is insufficient to account for the insulating gap of the system.

\begin{figure*}[!t]
\centering
\includegraphics[width= 17 cm]{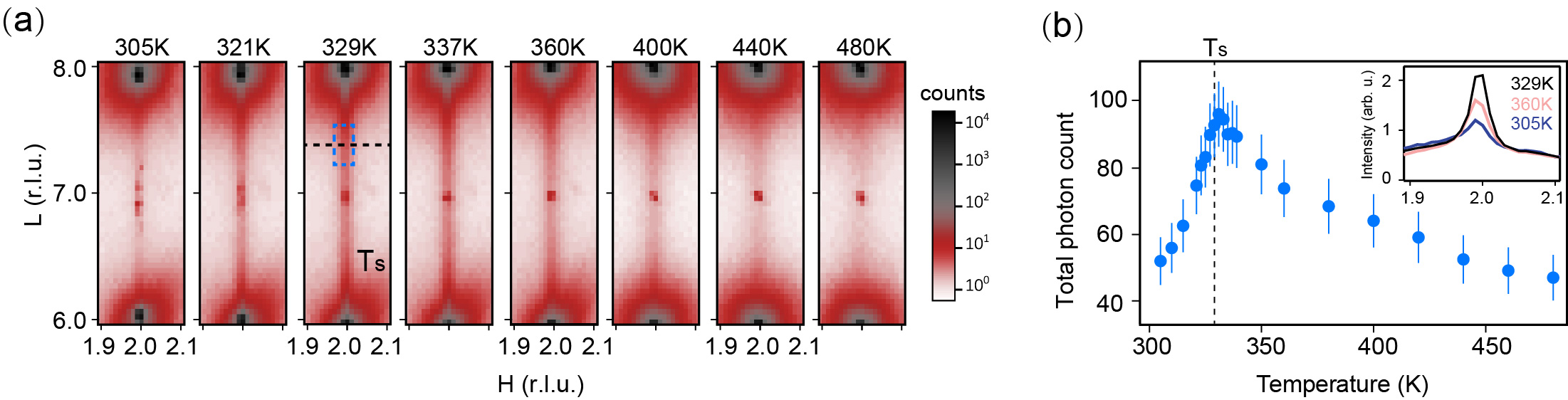}\vspace{-3mm}
\caption{Strong dynamic lattice fluctuation in \TNS\ above $T_{s}$. (a) Temperature dependence of the diffuse scattering signal along the \textit{c*}-direction between the 2 0 6 and 2 0 8 Bragg peaks. All figures are plotted under the same log color scale to highlight the evolution. (b) Integrated intensity of the diffuse signal from the blue box in (a). Inset: line cuts [black dashed line in (a)] of the diffuse signal at $T$ = 305 K, 329 K, and 360 K, respectively. 
}
\label{fig:Fig2}
\end{figure*}

The organization of this paper is as follows. We report the lattice fluctuations evidenced from XRD and the pseudogap state measured by ARPES in the high-temperature phase in Secs.~\ref{sec:xray} and \ref{sec:ARPES} respectively. Next, we discuss the non-thermal tuning in the low-temperature phase, the emergence of NEC, and bound the Coulomb interaction strength in Sec.~\ref{sec:lowTemp} and Sec.~\ref{sec:Coulomb}. We then construct a minimal lattice model, estimate the EPC vertex strength, and explain the experimental spectra using the first-principles and many-body simulations in Sec.~\ref{sec:theory}. The combined impacts of Coulomb interactions and electron-phonon couplings are discussed in Sec.~\ref{sec:discussion}. Finally, we conclude by discussing the implication of this framework to other quantum materials with intertwined orders, and make a few predictions for the \TNS~system in Sec.~\ref{sec:conclusion}.

\section{High-Temperature Lattice Fluctuations}\label{sec:xray}

High-quality \TNS\ single crystals were grown via the chemical vapor transfer method with iodine (I$_2$) as the transport agent\,\cite{di1986physical,lu2017zero,baldini2020spontaneous}. Starting materials, composed of Ta powder (99.99\%), Ni powder (99.99\%), Se powder (99.99\%) with a nominal molar ratio 2:1:5, were fully ground and mixed inside the glovebox. An additional 50\,mg of iodine was then added to the mixed powder before it was vacuumed, backfilled with 1/3 Argon, and sealed inside a quartz tube (inner diameter 8\,mm, outer diameter 12\,mm, and a length of 120\,mm). The sealed quartz tube was placed horizontally inside a muffle furnace during the growth. The hot end reaction temperature was set to 950\,$ ^{\circ}$C and the cold end was left in air with the temperature stabilized at 850\,$^{\circ}$C. Long and thin single crystals were harvested by quenching the furnace in air after one week of reaction. Residue iodine on the surface of the crystals was removed with ethanol.

\TNS\ crystallizes in a layered structure stacked via van der Waals interactions [Fig.~\ref{fig:Fig1}(a)]. Within each layer, the Ta and Ni atoms form a chain structure along the \textit{a}-axis of the crystal. Our XRD data reveals that the system undergoes a structural phase transition from a high-temperature orthorhombic \textit{Cmcm} phase to a low-temperature monoclinic \textit{C2/c} phase at $T_s$, in line with the previous report\,\cite{di1986physical}. During the transition, the Ta atoms slightly shear along the chain direction, resulting in an increase of the {$\beta$} angle from $90^\circ $ to $90.53^\circ$ [exaggerated in Fig.~\ref{fig:Fig1}(a)]. The second-order nature of this transition is indicated by the continuous separation of the $\mathrm{\pm}$ 1 3 4 and $\mathrm{\pm}$ 1 3 7 nuclear Bragg peaks, which are identical above $T_s$ [Fig.~\ref{fig:Fig1}(b)]. Here, the quantity $|\beta-90^\circ|$, or equivalently the average displacement $x$ of the Ta atoms, is used as the structural order parameter of this phase transition. We further compare the width of the Bragg peaks below and above $T_s$ [Fig.~\ref{fig:Fig1}(c)]. The peak width is nearly identical across the structural transition, ruling out any vestigial static monoclinic phase in the high-temperature normal state.

In addition to the immediate nullification of the lattice order parameter above $T_s$, we also observe a pronounced decrease of the 2 0 0 nuclear Bragg peak intensity [Fig.~\ref{fig:Fig1}(c)], indicating the presence of strong lattice fluctuations. Specifically, this fluctuation originates from the squared average of the atomic displacement $x$, and is further corroborated by the dramatically enhanced \textit{c*}-direction diffuse scattering signal over a broad temperature range above $T_s$ [Figs.~\ref{fig:Fig2}(a)-(b)]. Such a diffuse scattering signal can occur when the atomic position deviates from perfect periodicity along the \textit{c*}-direction. Since static lattice disorder is ruled out, a strong dynamic lattice fluctuation in the form of inter-chain sheering is the most plausible scenario. This aligns with previous reports of the soft transverse acoustic phonon\,\cite{nakano2018antiferroelectric} and the electronically coupled optical phonons in \TNS\,\cite{yan2019strong,volkov2021critical,kim2021direct}.

\section{High-Temperature Pseudogap}\label{sec:ARPES}

\begin{figure*}[!th]
\centering
\includegraphics[width= 17 cm]{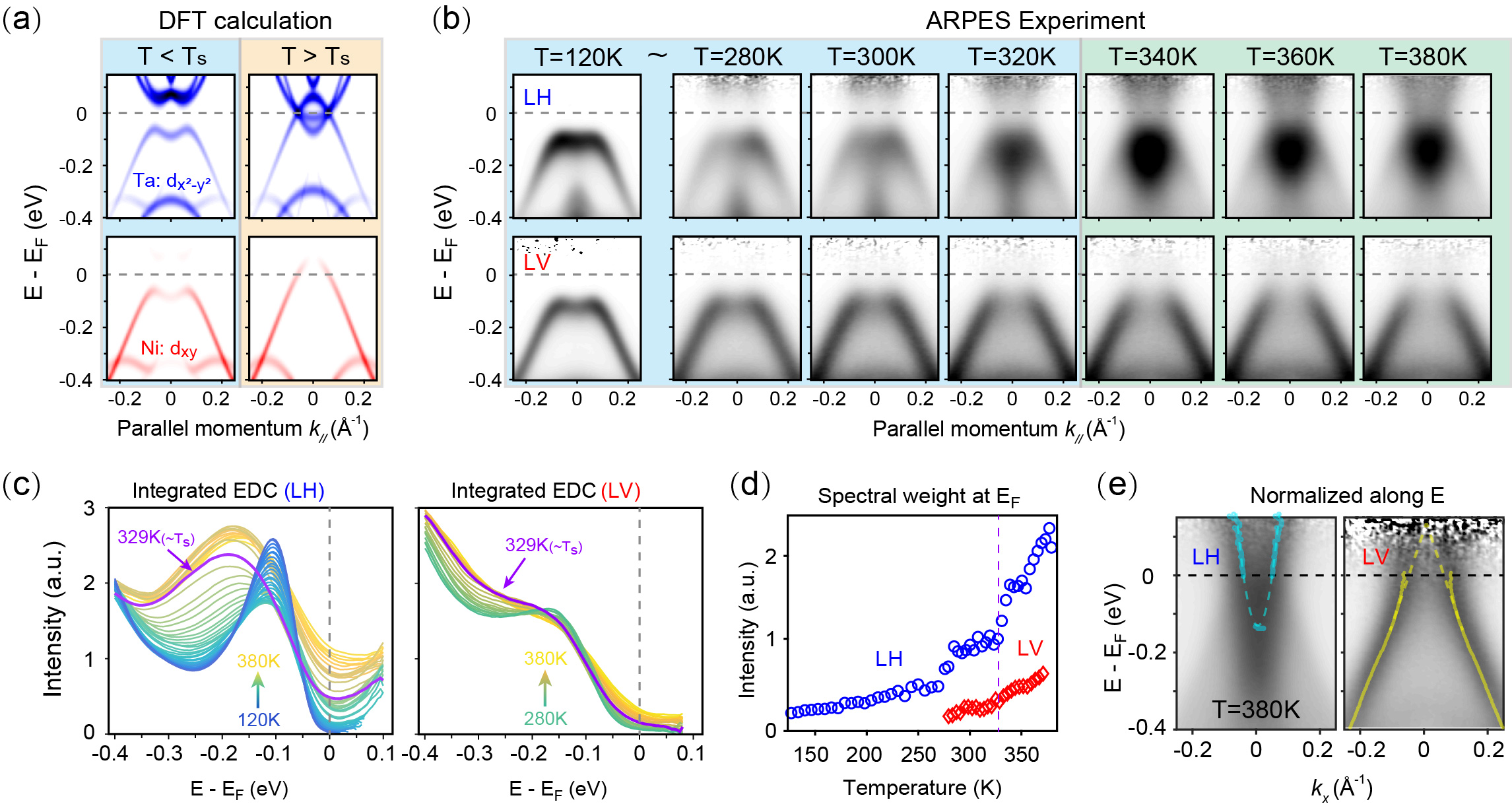}\vspace{-3mm}
\caption{Temperature dependence of the single-particle gap in \TNS. (a) DFT calculations of the single-particle bands below and above $T_{s}$ using a Fermi-Dirac smearing. States from Ta $d_{x^2-y^2}$ and Ni $d_{xy}$ orbitals are emphasized, corresponding to the linear horizontal (LH, blue) and linear vertical (LV, red) channels in the ARPES experiments, respectively. (b) Photoemission spectra along X-$\Gamma$-X direction in both photon polarization channels at select temperatures. (c) Temperature-dependent integrated energy distribution curves (EDCs) from the spectra in (b) for both photon polarization channels. (d) Temperature dependence of the integrated spectral weight around $E_F \mathrm{\pm}$ 30 meV. The vertical purple dashed line marks the $T_{s}$.  (e) Photoemission spectra at $T$ = 380 K, with intensity normalized along the energy axis to highlight the underlying dispersion.
}
\label{fig:Fig3}
\end{figure*}

The structural phase transition at $T_s$ could potentially drive an MIT transition in the electronic channel, taking the route of band folding in mean-field theory, which is discussed in many charge density wave (CDW) materials\,\cite{borisenko2008pseudogap,chen2015charge}. This synchronization of structural symmetry breaking and electronic MIT is also expected according to the first principles calculation of the electronic structure of \TNS~ based on density functional theory (DFT). As shown in Fig.~\ref{fig:Fig3}(a), the relaxed ground-state structure exhibits a lattice order parameter {$\beta$} comparable to the experimental data (detailed in Sec.~\ref{sec:theory}), and the calculated band structure exhibits an energy gap. When the structure is constrained to maintain orthorhombicity, the system is a gapless semimetal. However, previous transport studies have indicated only a small change of the resistivity slope across the structural transition at $T_s$, with an insulating behavior persisting up to 550\,K, more than 60\% above $T_s$\,\cite{di1986physical}.

To reveal the origin of this asynchronized electronic MIT and structural symmetry breaking, we investigate the single-particle spectral function of \TNS\ with high-resolution ARPES. Figure \ref{fig:Fig3}(b) presents the APRES data along the $\Gamma$-X high symmetry direction [geometry illustrated in Fig.~S1] for various temperatures below and above $T_s$. Here, the high statistics and energy resolution enable spectra restoration up to $\sim$150\,meV above the Fermi level ($E_F$) after dividing the resolution-convolved Fermi-Dirac function\,\cite{he2021superconducting}. As illustrated in Fig.~\ref{fig:Fig3}(b), the low-temperature spectra ($T < T_s$) show a pronounced single-particle gap, where the dispersion and orbital compositions are consistent with the DFT calculations in Fig.~\ref{fig:Fig3}(a). Therefore, the electronic state below $T_s$ aligns with the mean-field theory for a second-order phase transition.

In contrast, the high-temperature electronic structure ($T>T_s$) exhibits anomalous behavior beyond the DFT prediction. We observe a pronounced spectral weight depletion around $E_F$ ($\mathrm{\pm}$ 100\,meV), despite the disappearance of the low-temperature flat valence band top [Fig.~\ref{fig:Fig3}(b)]. Such a strong intensity depletion cannot be addressed in the generic band theory: drastic orbital character change alone cannot account for the missing spectral weight in both Ta and Ni states probed in both photon polarization channels. The dipole transition matrix-element effect is also unlikely, given the robust spectral weight depletion seen across a broad range of photon energies and Brillouin zones [Fig.~S2].

The relative evolution of this spectral depletion to the structural phase transition can be better illustrated from the integrated energy distribution curves (EDCs) [Fig.~\ref{fig:Fig3}(c)]. The structure-related band reconstruction at high binding energy (marked by the shifting of the valence band from -0.11\,eV to -0.17\,eV) happens within a relatively small temperature range below $T_s$\, in line with the second-order phase transition. In contrast, the low-energy spectral weight ($E_F\pm30$\,meV), drops continuously from 380 K to 120 K with little sensitivity to the structural phase transition [Fig.~\ref{fig:Fig3}(d)]. Such persistent single-particle spectral depletion at $E_F$ naturally accounts for the insulating behavior in resistivity and optical conductivity above $T_s$\,\cite{di1986physical,larkin2017giant}. 

Despite this spectral weight depletion, however, the band dispersions above $T_s$ do not suffer from apparent gap opening or band hybridization. This can be seen from the band dispersions extracted from the ARPES spectra, normalized along the energy axis to remove the intensity depletion effect. As shown in Fig.~\ref{fig:Fig3}(e), we identify uninterrupted conduction and valence band dispersions (recovered up to $\mathrm{\sim} 5k_BT_s$ above $E_F$) consistent with the DFT calculation. Such a single-particle ``gapped'' spectrum on top of a metallic dispersion, without a global symmetry breaking of the system, is analogous to the pseudogap state found in high-$T_c$ cuprates~\cite{alloul198989,timusk1999pseudogap,sobota2021angle}, and is recently shown to also relate to electron-hole excitations~\cite{singh2022unconventional}. 

Summing up the experimental results of \TNS\ upon thermal tuning, an electronic gapped state is found below $T_s$ in line with the second-order structure phase transition and the DFT calculation. Meanwhile, we observe concomitant evolutions of strong lattice fluctuation and electronic pseudogap state above $T_s$, without any global symmetry breaking, suggesting potential involvement of strong EPC behind the MIT transition in \TNS.

\begin{figure*}[!th]
\centering
\includegraphics[width= 17 cm]{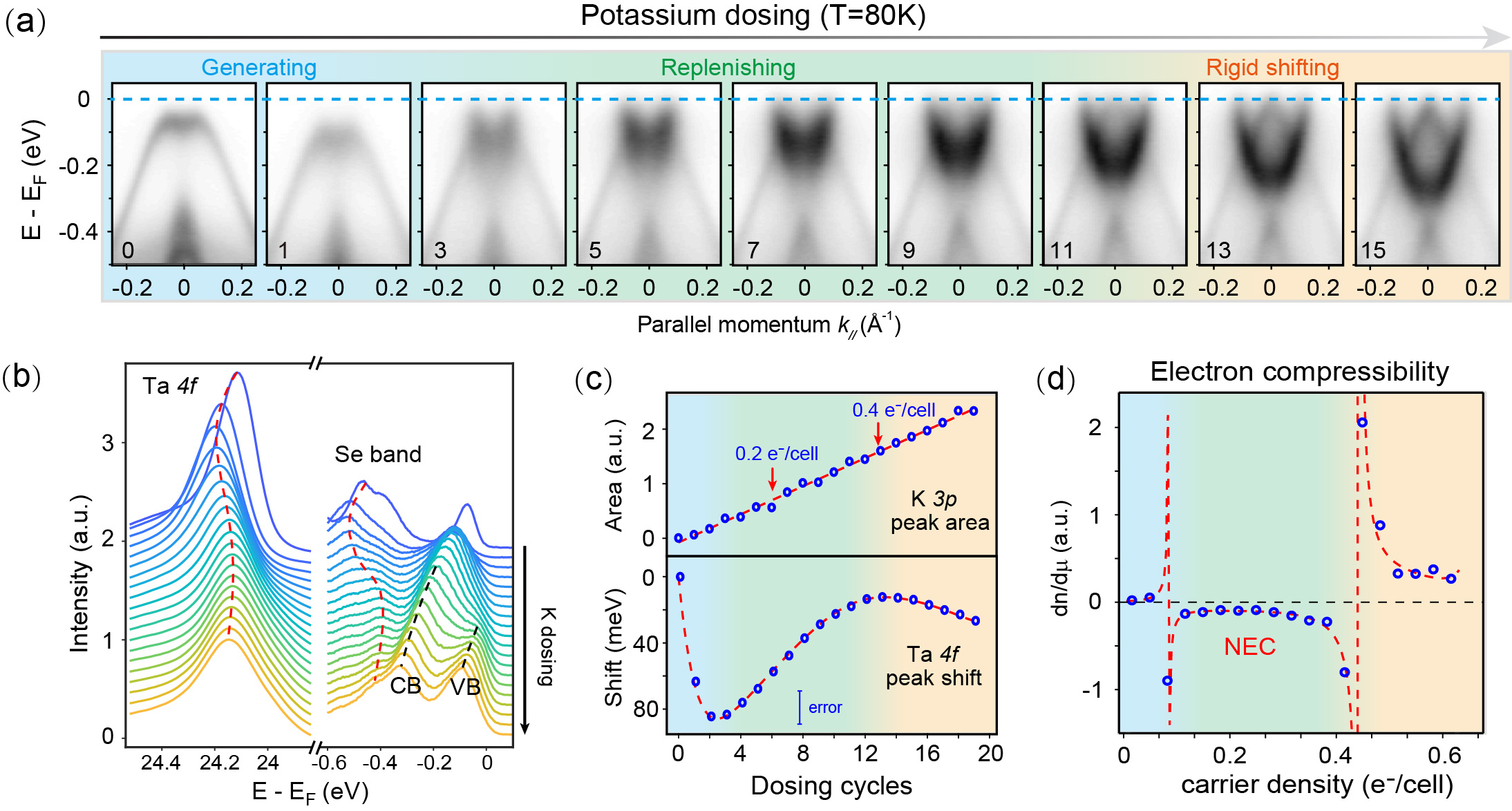}\vspace{-3mm}
\caption{Evidence of pseudogap and negative electronic compressibility (NEC) with potassium dosing. (a) Photoemission spectra along X-$\Gamma$-X directions at discrete potassium dosing cycles (lower-left indices) at 80K. Distinct regions are shaded in different colors, marking the evolution stages of the band structure. (b) EDCs at $\Gamma$ for both Ta $4f$ corelevel and low energy bands. The peak shifts with potassium dosing are illustrated with dashed lines. CB: Conduction Band. VB: Valence Band.  (c) Carrier density and chemical potential estimated from the peak area increment of the K $3p$ corelevel (Fig.~S3) and the peak position shift of the Ta $4f$ core-level in (c), respectively. (d) Electronic compressibility calculated from the parameters deduced in (d). Red line is calculated from the polynomial fit of the scattered data points in (d).
}
\label{fig:Fig4}
\end{figure*}

\section{Low-Temperature Pseudogap and Negative Electronic Compressibility}\label{sec:lowTemp}

While the concomitant evolution of the high-temperature electronic pseudogap and strong lattice fluctuation suggest a connected microscopic mechanism, one cannot quantitatively delineate the contributions of EPC and direct electron-hole Coulomb interaction (excitonic instability) to the broken-symmetry ground state. Non-thermal tuning methods offer more possibilities to separate the above two channels. While pump-probe studies could access excited states beyond the temperature range in equilibrium ARPES measurements, the non-equilibrium dynamics involve non-equilibrated electronic and lattice temperatures, as well as other non-thermal effects\,\cite{de2021colloquium}. Therefore, contentions remain as to which interaction term determines the eventual restoration of metallicity in \TNS\,\cite{mor2017ultrafast,okazaki2018photo,tang2020non,baldini2020spontaneous}.

Tuning the carrier density proves to be another efficient non-thermal tuning method in \TNS, which can restore the gapless state at low temperatures\,\cite{fukutani2019electrical,chen2020doping}. Here, through the combined use of a micron-spot synchrotron ARPES and \textit{in-situ} potassium (K) dosing in the low-temperature broken-symmetry state, we achieve a quantitative measure of both the charge doping and chemical potential shift [see Fig.~\ref{fig:Fig4}]. As will be discussed below, this measurement is necessary to examine the thermodynamic stability of the electronic subsystem. The relative potassium dosage is accessed through the core-level x-ray photoemission spectroscopy (XPS) of the K $3p$ orbital, and the electron doping per unit cell is determined by the Luttinger volume at different dosing levels (see Appendix \ref{app:doping} for details). 

Enabled by these improvements, our experiments reveal a three-stage isothermal spectral evolution across the dosing-tuned MIT [Fig.~\ref{fig:Fig4}(a) and Fig.~S3]. This evolution takes a striking resemblance to the thermal phase transition (insulator-pseudogap-semimetal). Starting from a low-temperature gapped state, the valence band rapidly downshifts upon dosing (shaded in blue). A pseudogap concurrently appears at $E_F$, similar to the spectral features right above $T_s$. In the second stage (shaded in green), the energy position of the conduction band barely changes while the missing spectral weight in the pseudogap is gradually replenished, resembling the pseudogap filling at higher temperatures. Finally, a semi-metallic state is restored with the full conduction and valence bands intersecting each other (shaded in orange). This semi-metallic state achieved with heavy potassium dosing is qualitatively identical to the normal state of \TNS\ in the high-temperature orthorhombic phase, as suggested by the pump-probe experiments\,\cite{tang2020non} and the DFT calculation [Fig.~\ref{fig:Fig3}(a)]. Such resemblance indicates that upon electron doping, the system may have a similar tendency to naturally recover the orthorhombic lattice structure, which is indeed confirmed by our first-principles calculation as will be detailed in Sec.~\ref{sec:theory}.

Meanwhile, thermodynamic inevitability of the lattice's participation is evidenced by the anomalous chemical potential evolution during the electron injection process. Here, the chemical potential change is evaluated from the energy shift of the fully filled Ta $4f$ core-level and cross-referenced with the Se valence band shift [Fig.~\ref{fig:Fig4}(b)], to rule out extrinsic contributions\,\cite{wen20203d}. The dosage-independent linewidth indicates insignificant inelastic scattering from potassium disorder. Both the Ta $4f$ core-level and the Se valence band are found to evolve non-monotonically with charge doping, in sharp contrast to the non-interacting scenario where a rigid shift of bands towards higher binding energy is expected. Such a non-monotonic behavior of the chemical potential can be described in terms of electronic compressibility. NEC, where the energy of the electronic subsystem decreases despite the addition of electrons, appears when the pseudogap is being replenished in the second stage of charge doping [the green region in Fig.~\ref{fig:Fig4}(c-d)]. Considering that purely electronic systems can only repel added electrons, such an NEC behavior requires the involvement of additional degrees of freedom, such as the lattice excitations, in order to maintain the thermodynamic stability of the full system. 

\section{Determination of the electron-hole Coulomb attraction}\label{sec:Coulomb}

\begin{figure}[!b]
\centering
\includegraphics[width= 8.5 cm]{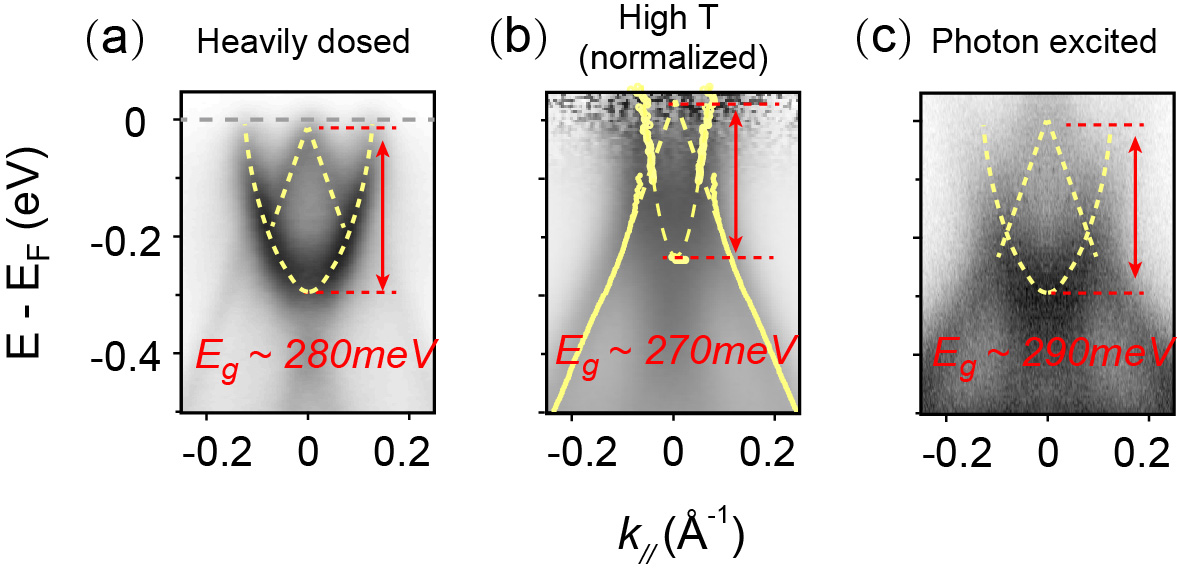}\vspace{-3mm}
\caption{Comparison of conduction and valence band overlap in the semi-metallic normal state of \TNS, achieved under different dielectric environments. (a) heavily potassium-dosed sample. (enhanced $n_e$) (b) high-temperature spectrum of the intrinsic sample, normalized along the energy axis. (intrinsic $n_e+n_p$) (c) photon-excited sample (enhanced $n_e+n_p$, reproduced from previous work\cite{tang2020non}). 
}
\label{fig:Fig5}
\end{figure}

\begin{figure*}[!tb]
\centering
\includegraphics[width= 17 cm]{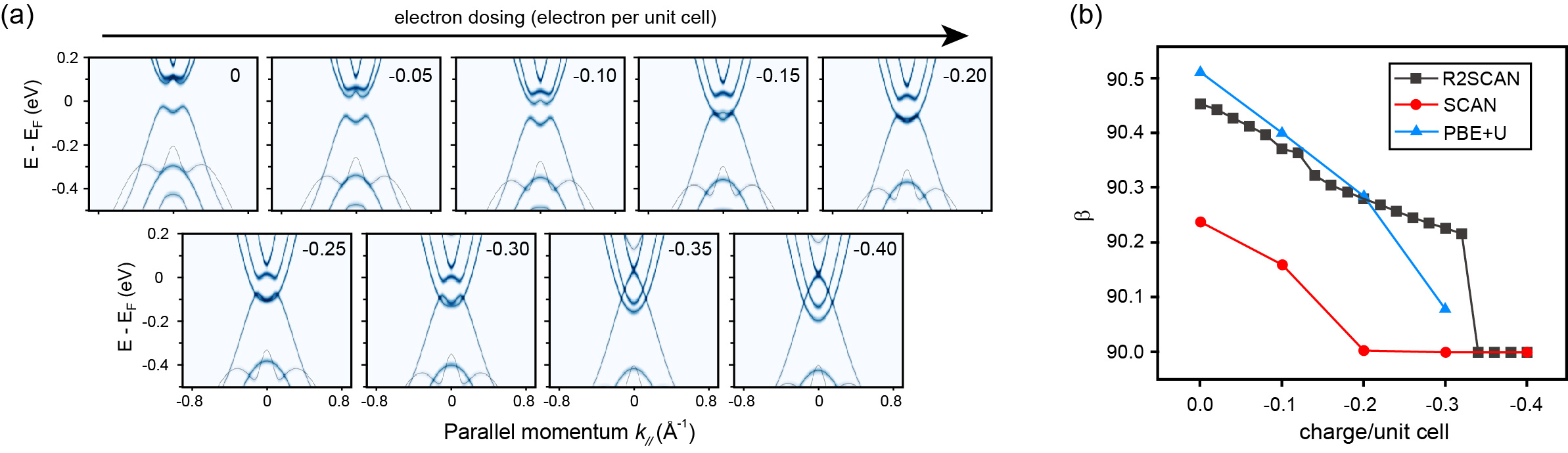}\vspace{-3mm}
\caption{Evolution of band structure with carrier doping by DFT calculation. (a) Evolution of the calculated band structure (using the r$^2$SCAN functional) with carrier doping starting from the low-temperature monoclinic fully gapped state. (b) Evolution of beta angle (marking the low-temperature monoclinic structure when deviated from $90^\circ $) evaluated with different DFT functionals.
}
\label{fig:Fig6}
\end{figure*}

An important consequence of accessing the semi-metallic normal state via different tuning methods is that it provides a reference to determine the Coulomb interaction strength, which are hard to estimate in the broken symmetry phase without bias. Here, we quantitatively compare the valence-conduction band overlaps $E_g$ in the semi-metallic normal state reached via heavy potassium dosing with the other two normal states achieved under dramatically different dielectric environments - equilibrium heating and neutral photodoping [see Fig.~\ref{fig:Fig5}]. To study the influence of direct inter-band Coulomb interaction on this $E_g$, we consider a three-orbital model when lattice effects are not included. The inter-band Coulomb interaction term reads as 
\begin{equation}\label{eq:threeBandInteraction}
    \mathcal{H}_{\mathrm{int}}=  V\!\sum_{i,\sigma ,\alpha ,\sigma^\prime} \left(n^c_{i\alpha \sigma}+n^c_{i+1\alpha \sigma}\right)n^f_{i\sigma^\prime}
\end{equation}
where $V$ is the Coulomb interaction strength. $n^c_{i}$, $n^c_{i+1}$, and $n^f_{i}$ are the density operators of two degenerate conduction bands (Ta $5d$) and the valence band (Ni $3d$).  If we compare normal states only, the system is resistant to exciton condensation (detailed in Sec.~\ref{sec:theory}) and each band has conserved electron numbers (for specific global doping). We denote their total electron numbers (considering the conduction-band degeneracy) as $N_c$ and $N_f$, respectively. Thus, the effective Hamiltonian can be mapped into 
\begin{equation}\label{eq:bandShiftc}
H^c_{\mathrm{int}}\approx V\sum_{i,\sigma ,\alpha ,\sigma^\prime} \left(n^c_{i\alpha \sigma}+n^c_{i+1\alpha \sigma}\right) \left\langle n^f_{i\sigma^\prime}\right\rangle =\frac{2VN_f}{N} \sum_{i,\alpha ,\sigma} n^c_{i\alpha \sigma}
\end{equation}
for the conduction band and
\begin{equation}\label{eq:bandShiftf}
H^f_{\mathrm{int}}\approx \ V\sum_{i,\sigma ,\alpha ,\sigma^\prime} \langle n^c_{i\alpha \sigma}+n^c_{i+1\alpha \sigma}\rangle n^f_{i\sigma^\prime}=\frac{2VN_c}{N}\sum_{i,\sigma} n^f_{i\sigma}
\end{equation}
for the valence band, where $N$ is the number of unit cells. As a result, in addition to excitonic instability, which disappears in those normal states, the Hartree part of the Coulomb interaction can be directly read out from the relative shift of both conduction and valence band site energies. The doping-induced change of the band overlap $E_g$ corresponds to $\mathrm{\Delta }\mu ={2V(\mathrm{\Delta }N_c-{\mathrm{\Delta }N}_f)}/{N}$. Therefore, the $E_g$ is found to be linearly reduced with increasing $V$ [see Fig.~S4]. Here $(\mathrm{\Delta }N_c-{\mathrm{\Delta }N}_f)/N$ can be estimated from the potassium dosed ARPES spectrum. According to the Luttinger theorem [Appendix \ref{app:doping}], the doped electron density for the 15$^{th}$-dosing [Fig.~\ref{fig:Fig4}(a)] reaches $\sim$0.4 per unit cell, indicating $(\mathrm{\Delta }N_c+{\mathrm{\Delta }N}_f)/N\ =0.46$. As the Fermi velocity is comparable in the valence and two conduction bands, it leads to $(\mathrm{\Delta }N_c-{\mathrm{\Delta }N}_f)/N\ \approx 0.152$. On the other side, $\mathrm{\Delta }\mu$ can be estimated by comparing the change of band overlap $E_g$ in the normal state [Fig.~\ref{fig:Fig5}]. The heavily K-dosed spectra ($E_g\cong 280$\,meV, heavily screened situation with enhanced \textit{$n_e$}), the high-temperature spectra ($E_g\cong 270$\,meV, intrinsic $n_e$ + $n_p$), and the pump-probe photodoped spectra\,\cite{tang2020non} ($E_g\cong 290$\,meV, photodoping enhanced $n_e$ + $n_p$) give an upper bound to the change of $E_g$ ($\mathrm{\Delta }\mu$) of $\sim 20$meV. Therefore, we can stringently place an upper bound of direct Coulomb interaction strength $V\sim 70$\,meV in the system. This Coulomb interaction, in stark contrast to the $\sim$ eV scale strength required in previous Coulomb-only mean field calculations\,\cite{mazza2020nature,seki2014excitonic}, is not nearly enough to account for the experimentally observed low-temperature gap by itself [further discussed in Fig.~\ref{fig:Fig9}].

Summing up our results on low-temperature charge doping, a semi-metallic electronic structure is recovered with heavy electron injection through a pseudogap stage, resembling the normal state reached at sufficiently high temperatures. Moreover, the thermodynamic inevitability of the lattice's participation is revealed through the appearance of the pseudogap state and the NEC, while the direct Coulomb interaction strength is estimated to be insufficient to account for the insulating state by itself. 

\section{Theoretical Simulations and Interpretations}\label{sec:theory}

The stringent upper bound for the allowed Coulomb interaction and the indispensable contribution from the lattice, revealed by the above experimental data, point to the need to reexamine the proposal of excitonic condensate formation in \TNS. Exciton condensation was first proposed as a viable route toward MIT soon after the success of the BCS theory of superconductivity\,\cite{jerome1967excitonic}. In \TNS, one leading contention is whether it is the electron-phonon coupling or direct electron-hole Coulomb attraction that ``dominates'' the transition. Based on the new experimental evidence discussed in Sec.~\ref{sec:xray}-\ref{sec:lowTemp}, we quantify the EPC and Coulomb terms, and systematically examine the roles of each term using first-principles and many-body methods in this section.

We first focus on the mechanism of the semi-metallic normal state restoration by charge doping in Fig.~\ref{fig:Fig4}(a). To describe this process, we simulate the system at different doping levels (see Appendix \ref{app:theoryDetails} for details), using first-principles DFT method with the numerically stable r$^{2}$SCAN functional\,\cite{furness2020accurate}. This functional includes the kinetic energy density contribution, whose accuracy in correlated materials is beyond the widely used standard functionals within local density approximation (LDA) or generalized gradient approximation (GGA)~\cite{pokharel2022sensitivity}. To reveal the ground-state structures, we perform lattice relaxation at each doping level to achieve the most energetically favorable lattice structure. As illustrated in Fig.~\ref{fig:Fig6}(a), starting from the low-temperature monoclinic gapped state, the system gradually restores semi-metallicity with a transition at the doping level of $x_{\text{doping}}\sim0.15$ (extra electrons per unit cell). Such a dramatic change in the electronic bands is mainly caused by the change of the lattice structure as shown in Fig.~\ref{fig:Fig6}(b). We follow the definition in Sec.~\ref{sec:xray} and use the $\beta$ angle to depict the structural transition. Cross-checked with different DFT functionals, $\beta$ decreases as additional electrons dope into the system and eventually reaches 90$\mathrm{^\circ}$ at $x_{\text{doping}}\sim 0.2-0.3$. This simulated doping evolution reflects that the lattice structure of \TNS\ has a natural tendency to reshape into the high-symmetry orthorhombic phase (the high-temperature lattice structure in Sec.~\ref{sec:xray}) upon charge doping. It is worth noting that DFT is a ground-state theory and the functionals we adopted do \textit{not} include excitonic effects. But the impact of phonons has been included on the level of Born-Oppenheimer approximation through the structural optimization at each doping level. Therefore, the experimentally observed restoration of the high-symmetry lattice structure upon charge doping (see Sec.~\ref{sec:lowTemp}), can be explained as a concomitant effect caused by changes in the lattice structure, without invoking the influence of additional excitonic instability. 

To quantify the impact of phonons, we further evaluate the electron-phonon coupling strength near the Fermi level. Following Ref.~\onlinecite{giustino2017electronphonon}, the coupling matrix element among Kohn-Sham orbitals is
\begin{equation}\label{eq:el-ph}
g_{mn}^{(\nu)}(\boldsymbol{k},\boldsymbol{q}) = \langle u_{m \boldsymbol{k+q}}|\Delta^{(\nu)}_{\boldsymbol{q}}V^{\rm KS}|u_{n\boldsymbol{k}}\rangle \,,
\end{equation}
where $|u_{n\boldsymbol{k}}\rangle$ is the Bloch periodic part of the Kohn-Sham wavefunction with momentum $\boldsymbol{k}$ of the $n$-th band; the $\Delta^{(\nu)}_{\boldsymbol{q}}V^{\rm KS}$ is the variation of Kohn-Sham potential induced by the $\nu$-th phonon mode at momentum $\boldsymbol{q}$. The definition of the dimensionless operator $\Delta^{(\nu)}_{\boldsymbol{q}}$ and details of the simulation are described in Appendix \ref{app:theoryDetails}. 
 
The orthorhombic phase of Ta$_2$NiSe$_5$ is unstable at low temperature [see Figs.~\ref{fig:Fig3}(a-b) and Fig.~\ref{fig:Fig6}(a)]. Forcing the lattice structure to stay in a high-symmetry orthorhombic phase, our first-principles simulation identifies two unstable phonon modes responsible for the orthorhombic-monoclinic structural transition. However, one cannot directly evaluate the electron-phonon coupling matrix elements within the DFT framework for these unstable phonons. Instead, we employ the reference system Ta$_2$NiS$_5$ to estimate these couplings, whose orthorhombic phase is stable at low temperature. In Ta$_2$NiS$_5$, we identify two phonon modes, labeled by their irreducible representation $B_{2g}$ and $B_{1g}$, that have the maximum eigenvector overlap with the two unstable modes in \TNS, respectively. In Ta$_2$NiS$_5$, these two phonon modes couple to three Kohn-Sham bands near Fermi surface: the top valence band ($v_1$), the lowest conduction band ($c_1$), and the second lowest conduction band ($c_2$). The coupling matrix elements (at the long-wavelength phonon limit $q=0$) in this three-orbital subspace are
\begin{eqnarray}\label{eq:DFTEPC}
\left|g_{v_1c_1}^{(B_{2g})}(\boldsymbol{k}=\Gamma,\boldsymbol{q}=0)\right|&=&52.4\,\text{meV}
\nonumber\\
\left|g_{v_1c_2}^{(B_{1g})}(\boldsymbol{k}=\Gamma,\boldsymbol{q}=0)\right| &= &50.8\,\text{meV}\,.
\end{eqnarray}
The calculated phonon frequencies at $\boldsymbol{q}=0$ are $\omega_{B_{2g}}=1.91$\,THz and $\omega_{B_{1g}}=1.54$\,THz, consistent with those identified in \TNS~ by Raman spectra \cite{yan2019strong,volkov2021critical,kim2021direct}. Therefore, the phonon parameters extracted here will be used for further analysis in \TNS.

\begin{figure}[!t]
\centering
\includegraphics[width= \columnwidth]{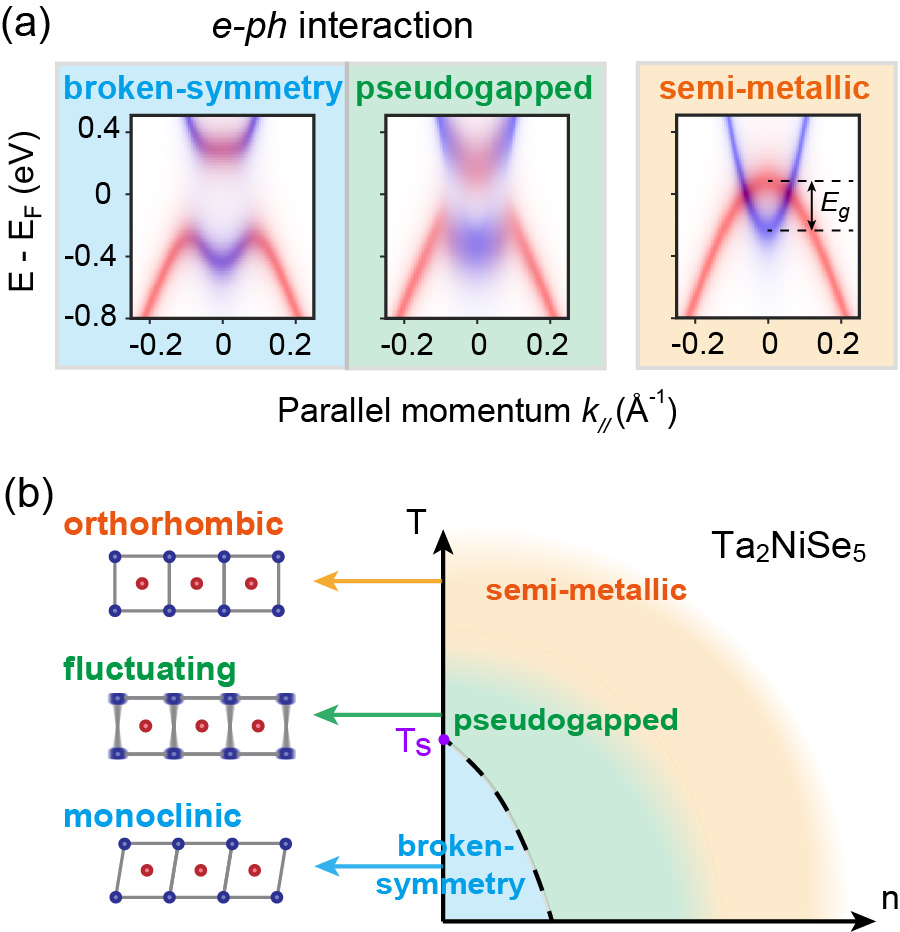}\vspace{-2mm}
\caption{Lattice fluctuation induced pseudogap and the temperature-doping phase diagram of \TNS. (a) Simulated band structure from the two-band model with only interband electron-phonon coupling ($g=70$\,meV): the broken-symmetry state caused by structural band hybridization (blue), the pseudogap state caused by lattice fluctuation (green), and the semi-metallic normal state (orange). (b) Temperature-doping phase diagram of \TNS. The crystal structures within different regions are illustrated on the side.
}
\label{fig:Fig7}
\end{figure} 

As we will show later, simply considering electron-phonon coupling at the mean-field phonon level (Born-Oppenheimer approximation) cannot address the asynchronization between the structural phase transition and electronic spectral evolution along the temperature axis, discussed in Sec.~\ref{sec:ARPES}. Previous theoretical investigations have concluded that an eV-scale Coulomb interaction is required to break the symmetry\,\cite{mazza2020nature,seki2014excitonic}, also without considering the non-mean-field phonon effects. With the parameters estimated from the experimental data and the first-principles simulations, we consider a many-body model explicitly including both electronic interactions and electron-phonon coupling and all orders of fluctuations. Here, we employ a model Hamiltonian in a similar form to Ref.~\onlinecite{kaneko2013orthorhombic}:
\begin{eqnarray}\label{eq:manyBodyHam}
\mathcal{H}\!\!&=&\!\! \mathcal{H}_0+\sum_{k,q,\sigma}\frac{g_q}{\sqrt{N}}\left[\left(a_q+a^\dagger_{-q}\right)c^\dagger_{k+q,\sigma}f_{k\sigma}+H.c.\right]\nonumber\\
&&+\sum_q{{\omega }_q\ a^\dagger_qa_q}
+V\sum_{i,\sigma ,\sigma^\prime}\left(n^c_{i\sigma}+n^c_{i+1\sigma}\right)n^f_{i\sigma^\prime}\,,\\
\mathcal{H}_0\!\!&=&\!\!\sum_{k,\sigma}{\varepsilon^c_kc^\dagger_{k\sigma}c_{k\sigma}}+ \sum_{k,\sigma}{\varepsilon^v_kf^\dagger_{k\sigma}f_{k\sigma}}\,,
\end{eqnarray}
but the band and interaction parameters are chosen to match the experimental data in this paper and the phonon parameters are chosen according to our first-principles simulations. We write the Hamiltonian in momentum space here: $c^\dagger_{k\sigma}$ ($c_{k\sigma}$) creates (annihilates) an electron at the conduction band (primarily Ta $5d$) for momentum $k$ and spin $\sigma$, with dispersion given by $\varepsilon^c_k$, while the  $f^\dagger_{k\sigma}$ ($f_{k\sigma}$) creates (annihilates) an electron at the valence band (primarily Ni $3d$), with dispersion given by $\varepsilon^v_k$. The $n^c_{i\sigma}$ and $n^f_{i\sigma}$ are the density operators for the conduction and valence band, respectively. The fitted band structures via experimental results [Fig.~\ref{fig:Fig3}(e)] read as
\begin{eqnarray}
&&\varepsilon^c_k=3.1 - 1.8\,{{\mathrm{cos}}}(k) - 0.9\,{{\mathrm{cos}}}(2k) - 0.6\,{{\mathrm{cos}}}(3k)\\
&&\varepsilon^v_k=-1.8 +1.5\,{{\mathrm{cos}}}(k) +0.3\,{{\mathrm{cos}}}(2k) +0.1\,{{\mathrm{cos}}}(3k)\,
\end{eqnarray}
The direct hybridization between the conduction and valence bands around $\Gamma$ is forbidden in the orthorhombic phase by the inversion symmetry\,\cite{mazza2020nature}. However, this hybridization is enabled by an $B_{2g}$ lattice distortion, parametrized as a momentum-dependent displacement $x_q^{(B2g)}$ and quantized as the phonon mode $x_q^{(B2g)}=a_q+a^\dagger_{-q}$.  As the Fermi momentum $k_F$ is much smaller than $2\pi /a_0$, we further restrict the e-ph coupling to the zone center $g_q=g{\delta }_q$. 
The inter-orbital Coulomb interaction term takes the same nearest-neighbor approximation as Eq.~\eqref{eq:threeBandInteraction}. Ideally, the conduction band should contain two sub-bands due to the two Ta chains per unit cell. However, we consider only the $B_{2g}$ coupling corresponding to the lower conduction band\,\cite{kaneko2013orthorhombic}, where the phonon mode reflects the strongest Raman anomaly across the transition\cite{yan2019strong,volkov2021critical,kim2021direct}. 

This model allows beyond-Born-Oppenheimer lattice fluctuations that couple to the electrons, and captures both the thermal and non-thermal evolution of the spectral function. We simulate single-particle spectral functions of the model in Eq.~\ref{eq:manyBodyHam} using the exact diagonalization (ED), which accurately describes all correlation effects in a finite system (see Appendix \ref{app:theoryDetails} for details). Here, $g$ is chosen as $70$\,meV, slightly larger than the number from the Ta$_2$NiS$_5$ simulation in Eq.~\eqref{eq:DFTEPC} to compensate the ignored electronic and lattice degrees of freedom, and we first exclude the Coulomb interaction by setting $V=0$ (see Sec.~\ref{sec:discussion} for the discussion about the impact of Coulomb interaction). Enhanced by the low dimensionality, strong lattice fluctuation above $T_s$ readily enables the conduction-valence band hybridization, and results in an electronic pseudogap without causing a global symmetry breaking [Fig.~\ref{fig:Fig7}(a), green]. Further reducing temperature results in a divergence of the phonon number, which signifies the phonon condensation and drives the transition to the monoclinic phase. Subsequently, a hard hybridization gap forms below $T_s$ [Fig.~\ref{fig:Fig7}(a), blue]. On the other hand, increasing electron carrier density lifts the Fermi level from the charge-neutral point and reduces the interband phonon dressing. Thus, doping ultimately drives the monoclinic phase into the orthorhombic phase even at low temperatures, which in turn reflects the first-principles results in Fig.~\ref{fig:Fig6}. The observed NEC also reflects strong electron-phonon coupling\,\cite{grilli1994electron,castellani1995singular} [see Appendix \ref{app:NEC}], where additional electronic energy from the added carriers is absorbed into the lattice degree of freedom. Therefore, the phase diagram of \TNS\  along both the temperature and electron-doping axes can be depicted in Fig.~\ref{fig:Fig7}(b), where the region shaded in green represents the lattice fluctuation induced pseudogap state above $T_s$. Clearly, both the thermal and quantum fluctuation effects are strong in this state.

When the phonon fluctuations are sufficient to drive the symmetry breaking through EPC, the many-body model in Eq.~\eqref{eq:manyBodyHam} can reproduce the sharp band gap of DFT simulations within the Born-Oppenheimer approximation. To show this, we simulate the spectral function by projecting the phonon fluctuations to a classical displacement (so-called the frozen phonon), which is equivalent to a coherent-state phonon wavefunction. Specifically, we assume that the ground state collapses to a broken-symmetry state characterized by a finite $X$. Treating this $X$ as a classical variable, we obtain the frozen-phonon equation for electrons
\begin{eqnarray}\label{eq:frozenPhonon}
\mathcal{H}_{\rm FP}\left(X\right)=\mathcal{H}_0+g_0X\sum_{k,\sigma}{\left[c^\dagger_{k\sigma}f_{k\sigma}+H.c.\right]}\ .
\end{eqnarray}
Using the $\langle (x_{q=0}^{\rm (B2g)})^2\rangle $ obtained by ED simulations for various temperatures, we estimate the frozen-phonon displacement by $X_{\rm ED}=\langle (x_{q=0}^{\rm (B2g)})^2\rangle^{1/2}$ and evaluate the spectral functions for the corresponding broken-symmetry states. As shown in the left panel of Fig.~\ref{fig:Fig7}(a), this mean-field treatment for phonons reproduces the gapped ARPES spectra with well-defined single-particle band folding. 

\section{Discussion}\label{sec:discussion}

\begin{figure}[!t]
\centering
\includegraphics[width= \columnwidth]{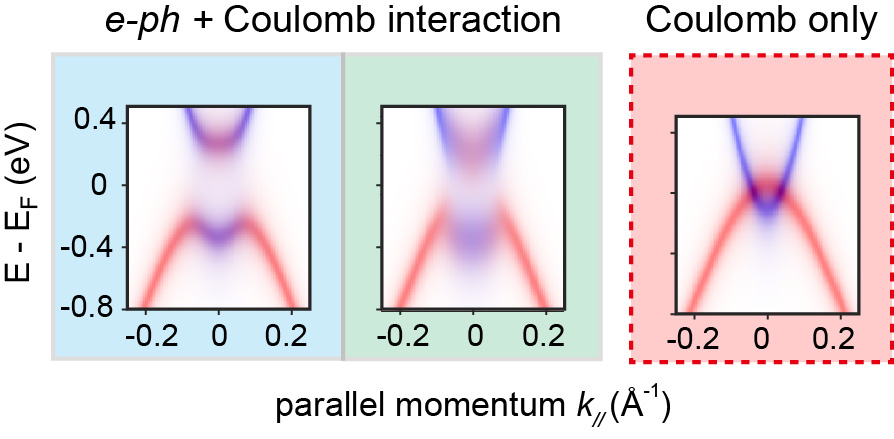}\vspace{-2mm}
\caption{Impact of Coulomb interactions. Blue and green: Simulated band structure with both EPC ($g=70$\,meV) and direct interband Coulomb interaction ($V=100$\,meV). Red: Simulated band structure with only interband Coulomb interaction ($V=100$\,meV). 
}
\label{fig:Fig8}
\end{figure}

We emphasize that our discussion of the EPC effects does not exclude the presence of electronic Coulomb interactions, which naturally exist in all materials. Instead, the quantification of a minimal lattice model provides a way to delineate the contributions from both interactions, avoiding the ``chicken-or-egg'' dilemma. We discuss the combined impact of EPC and electronic Coulomb interactions in this section. Figure \ref{fig:Fig8} shows the single-particle spectra with the EPC $g$ and interband Coulomb interaction $V$, whose upper bound is set by normal-state experimental results in Sec.~\ref{sec:lowTemp}. Comparing Fig.~\ref{fig:Fig7}(a) with Fig.~\ref{fig:Fig8}, we find that the Coulomb interaction within the allowed regime does not obviously influence the electronic structure, both in the broken-symmetry and pseudogap states. In contrast, when the EPC is ignored, the $\leq 100$\,meV Coulomb interaction can only repel the conduction and valence bands in their entirety at the long-wavelength limit [Fig.~\ref{fig:Fig8} dashed red box]. Such a relative band shift is not helpful -- and can even be destructive -- to interband hybridizations. 

\begin{figure}[!t]
\centering
\includegraphics[width= \columnwidth]{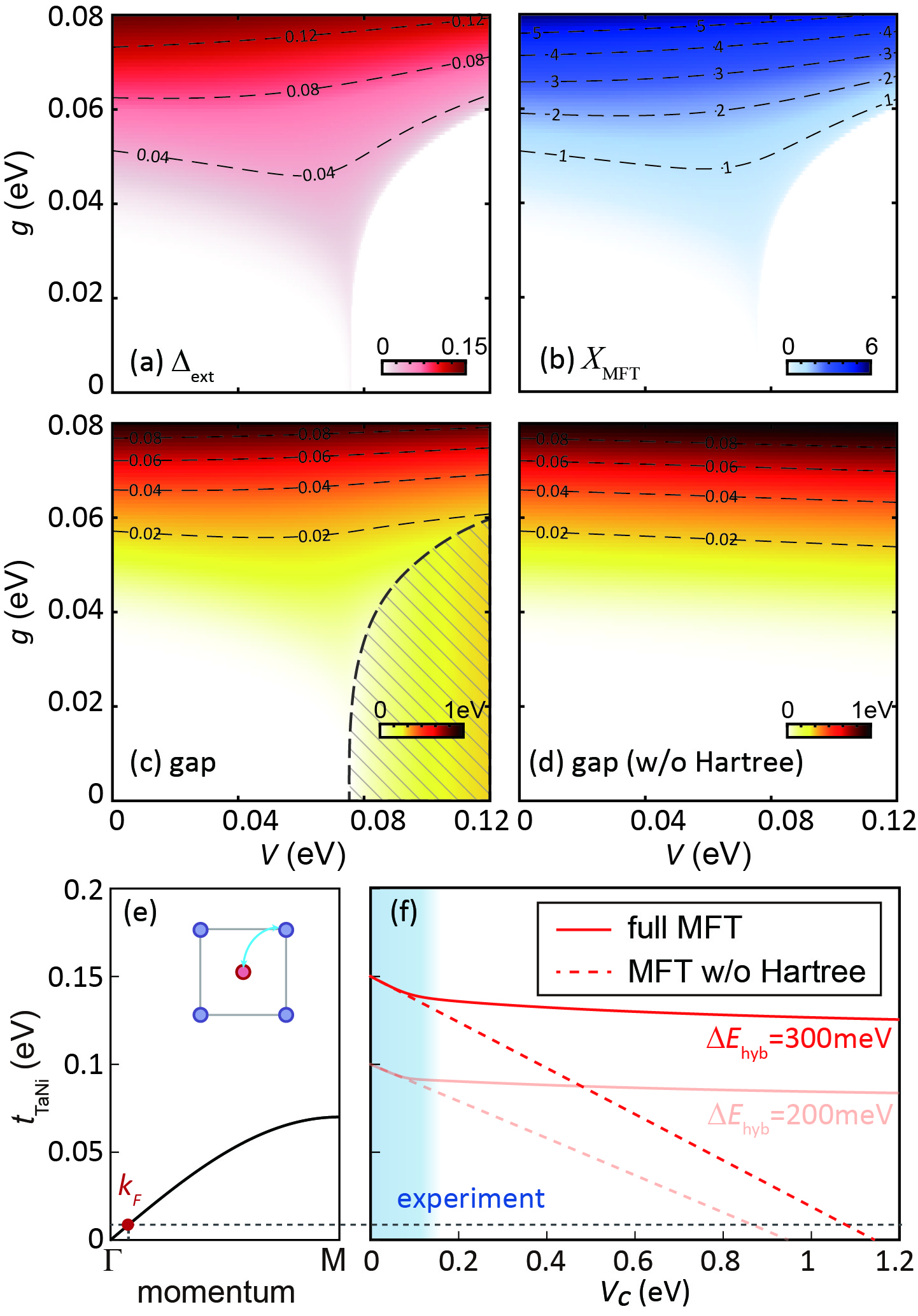}\vspace{-2mm}
\caption{Mean-field analysis of the \TNS\ model with the combined impact of EPC and interband Coulomb interaction. (a) The dependence of the excitonic order parameter $\Delta_{\rm ext}=\sum_{i\sigma}\langle f^\dagger_{i\sigma} c_{i\sigma}\rangle/2N$ on the EPC $g$ and Coulomb interaction $V$, obtained by the mean-field solution of Eq.~\eqref{eq:manyBodyHam}. (b) The same as (a) but for the lattice order parameter $X_{\rm MFT}$. (c, d) The dependence of the direct gap size on the two interactions for the (c) full mean-field solution and (d) solution without the Hartree term. The shaded regime indicates the opening of the band gap due to the relative shift of two bands, signaled by $\Delta_{\rm ext}=0$. (e) The strength of interband hopping integral $t_{\rm TaNi}$ as a function of momentum in the \TNS. (f) the critical Coulomb interaction strength $V_c$ needed to generate 300 and 200\,meV hybridization gaps $\Delta E_{\rm hyb}$ at given $t_{\rm TaNi}$. Solid line: the result of mean-field simulation. Dashed line: the result excluding the contribution from the Hartree part of the mean-field decomposition. The region shaded in blue: experimentally determined Coulomb interaction strength. 
}
\label{fig:Fig9}
\end{figure} 

More intuitions about the impact of EPC and Coulomb interactions can be obtained by analyzing the mean-field order parameters when both interactions are included. Thus, we further conduct a mean-field simulation for Hamiltonian in Eq.~\eqref{eq:manyBodyHam} with two order parameters: the electronic (excitonic) order parameter is defined as $\Delta_{\rm ext}=\sum_{i\sigma}\langle f^\dagger_{i\sigma} c_{i\sigma}\rangle/2N$, reflecting the hybridization between Ta to Ni orbitals; the lattice order parameter is $X_{\rm MFT}=\langle x_{q=0}^{\rm (B2g)}\rangle$, reflecting the symmetry breaking of the crystal and demonstrating the same information as $|\beta-90^\circ|$ in Sec.~\ref{sec:xray}\,\cite{phononDiff}.  As shown in Figs.~\ref{fig:Fig9}(a) and (b), these two order parameters follow the same dependence on interactions at the mean-field level. For weak interactions, the EPC ($g$) and Coulomb interaction ($V$) are cooperative in driving the hybridization $\langle f^\dagger c\rangle$ and forming excitons. However, since the Hartree part of the Coulomb interaction [i.e.~Eqs.~\eqref{eq:bandShiftc} and \eqref{eq:bandShiftf}] separates the two bands, it competes with the hybridization or exciton formation. When $V$ is larger than some critical value ($\sim 80$\,meV), its suppression on the excitonic or lattice order overwhelms the latter. 

To connect the order parameters with the single-particle spectra, we simulate the single-particle band gap within the mean-field framework. Note that these gaps are different from the ``pseudogap'' from Figs.~\ref{fig:Fig7} and \ref{fig:Fig8}, the latter of which contains many-body fluctuations beyond the mean field. Figure \ref{fig:Fig9}(c) shows the mean-field gap size for different combinations of both interactions. For weak Coulomb interaction, this gap is proportional to the excitonic order parameter $\Delta_{\rm ext}$ (and the lattice order parameter $X_{\rm MFT}$). They start to deviate from each other for a relatively strong $V$, denoted by the dashed line. In this case, the Hartree part of the Coulomb interaction dominates and a band gap opens, suppressing the hybridization. Obviously, overlapping conduction and valence bands are always observed in ARPES experiments in the normal state of \TNS, further supporting the conclusion about limited Coulomb interaction strength. To avoid the impact of this band gap, we further solve an artificial mean-field equation by eliminating the Hartree terms. This treatment leads to a pure hybridization gap [see Fig.~\ref{fig:Fig9}(d)], depending quadratically on $g$ and slightly on $V$. Detailed discussions about the hybridization gap and the impact of the Hartree terms are presented in Appendix \ref{app:hybridizationGap}.

The inefficiency of the interband Coulomb interaction in \TNS\ originates from the material's mirror symmetry with respect to the Ni-chain at small momenta, where the exact symmetry is present at $k = 0$\,\cite{mazza2020nature}. To illustrate this relation, we take the multi-band hopping parameters from Ref.~\onlinecite{mazza2020nature} and extract the component corresponding to the Ta $5d_{x^2-y^2}$ - Ni $3d_{xy}$ hopping integral, denoted as $t_{\rm TaNi}$. Due to the different parities, $t_{\rm TaNi}$ is an odd function of the momentum $k$ and its strength at the $k_F$ (determined by experiments) is extremely small compared to the bandwidth [see Fig.~\ref{fig:Fig9}(e)]. Adding this interband hopping $t_{\rm TaNi}$ in our two-band model, we conduct the same mean-field simulation for the order parameters and single-particle gap. Figure.~\ref{fig:Fig9}(f) shows the critical Coulomb interaction $V_c$ necessary to generate a 200-300\,meV hybridization gap ($\Delta E_{\rm hyb}$)\,\cite{hybGap}, without contributions from phonons. Due to the competition from the Hartree term, a hybridization gap comparable to experiments can not be reached given the experimentally bounded $V$. Excluding the contribution from the Hartree part of the mean-field decomposition can relax the requirement on $V_c$ to a certain extent (dashed line)\,\cite{seki2014excitonic}. However, compared to other excitonic insulator candidates, the required $V_c$ in \TNS\ is still far beyond the experimentally identified upper bound (blue shade), due to its uniquely small Fermi momenta (black dashed line).

\section{Conclusion}\label{sec:conclusion}
Our result reveals the important role of EPC in both the normal and broken-symmetry states of \TNS, based on the electrons' coupling to a strongly fluctuating lattice. Combining advanced spectroscopy and computation methods, we determine and quantify a minimal lattice model that captures the pseudogap state, the broken symmetry state, and the negative electronic compressibility. The direct electron-hole Coulomb interaction may not be zero but is not required to capture the above experimental observations. In the pseudogap phase without any global structural symmetry lowering, phonons cannot be treated under the Born-Oppenheimer approximation, under which the lattice's impact on the electronic structure is only reflected by a nonzero lattice distortion $\langle x\rangle $. Instead, phonons act on the electronic structure through a fluctuating state where $\langle x\rangle =0$ but ${\langle x}^2\rangle \ \neq \ 0$. In this regard, the pseudogap state may be conceptually likened to the `preformed excitons' proposed in recent studies\,\cite{fukutani2021detecting}, except that the binding is facilitated by interband electron-phonon coupling. In comparison, fluctuating CDW states in cuprates, K$_{0.3}$MoO$_{3}$, ZrTe$_{3}$, NbSe$_{3}$, and NbSe$_2$ are usually intertwined with electronic instabilities induced by strong Coulomb interaction\,\cite{sobota2021angle,pouget1985structural,schafer2001high,chatterjee2015emergence,yokoya2005role}. In \TNS, however, the major distinction lies in the small momentum low-energy band, which prohibits Coulomb interaction-induced charge transfer to the leading order. Such distinct lattice-symmetry-protected low-energy band crossings can be used to engineer accentuated lattice fluctuation effects in correlated materials.

The strong EPC in the \TNS~system also leads to predictions of distinct material properties upon further thermal and nonthermal tuning. First, in the strong-coupling scenario, the phase transition should not be tied to the Fermi surface topology. Recent studies showed S-doping as an effective method to control the conduction and valence band overlap in \TNS~\cite{lu2017zero}. The minimal many-body model shall produce an evolution of the phase transition temperature that is insensitive to the S-doping-tuned semimetal-to-semiconductor transition in Ta$_2$Ni(Se,S)$_5$, much different from the popular expectation of the dome-shaped excitonic insulator phase diagram upon band-gap tuning~\cite{jerome1967excitonic}. Second, given such strong EPC, the phonon self-energy will exhibit anomalies over a broad momentum range extending towards $\mathbf{q}\rightarrow0$, rather than a narrow region sharply centered around $\mathbf{q}\sim2\mathbf{k}_F$~\cite{zhu2015classification}. This can be verified in the dynamic structural factor accessed with high-resolution inelastic x-ray or neutron scattering experiments.

While we qualitatively explain the observed spectral behaviors using a single phonon mode, further experiments have indicated that multiple phonons are involved in this transition: the diffuse scattering and inelastic x-ray scattering results point to extensive transverse acoustic phonon softening\,\cite{nakano2018antiferroelectric}; Raman scattering also suggests strong involvement of electronically coupled optical phonons\,\cite{yan2019strong,volkov2021critical,kim2021direct}. Distinct from the acoustic-phonon-driven structural transition in 3D perovskite PrAlO$_{3}$\,\cite{birgeneau1974cooperative}, the strong fluctuations in \TNS\ seem to suggest additional contributions from the lower system dimension and the optical phonons. The strong lattice fluctuations and electron-phonon coupling in this system offer a knob to realize optically, electrically, and mechanically controlled MIT\,\cite{liu2021photoinduced}, and a platform to investigate the role of electron-phonon coupling behind phase instabilities in low dimensional systems.

Last but not least, we demonstrated a general method to unambiguously resolve the typical chick-or-egg problems in correlated material systems. First-principles and many-body numerical simulations can be combined with spectroscopic experiments not only to determine the minimal microscopic lattice model, but also to quantify the microscopic interaction parameters -- especially through numerous tuning methods. The minimal many-body model can be considered a faithful description of the material under investigation, and the roles of each interaction -- if a reductionist view is desired -- may be discussed by performing numerical experiments considering one term at a time. Most importantly, further material property predictions or engineering guidance can then be made by solving the perturbed minimal many-body model containing experimentally determined interaction parameters. This approach is ready to be generalized in low-dimensional correlated materials amid the rapid development of advanced spectroscopy, \textit{in situ} material tuning, and modern computational methods.

\section*{ACKNOWLEDGEMENTS}

We thank E. Baldini, S. D. Chen, E. W. Huang, A. Kemper, D. H. Lee, Y. B. Li, D. Y. Qiu, M. Trigo, Z. X. Shen, M. Yi, A. Zong, and W. T. Zhang for helpful discussions. Use of the Stanford Synchrotron Radiation Light Source, SLAC National Accelerator Laboratory, is supported by the US Department of Energy, Office of Science, Office of Basic Energy Sciences under Contract No. DE-AC02-76SF00515. This research used resources of the Advanced Light Source, a US DOE Office of Science User Facility under Contract No. DE-AC02-05CH11231. This work is based upon research conducted at the Center for High Energy X-ray Sciences (CHEXS) which is supported by the National Science Foundation under award DMR-1829070. Work at Lawrence Berkeley National Laboratory was funded by the U.S. Department of Energy, Office of Science, Office of Basic Energy Sciences, Materials Sciences and Engineering Division under Contract No. DE-AC02-05-CH11231 within the Quantum Materials Program (KC2202) and within the Theory of Materials Program (KC2301) which provided the DFT calculations. S.D. and Y.W. acknowledge support from the National Science Foundation (NSF) award DMR-2038011. The electron-phonon model simulations were performed on the Frontera computing system at the Texas Advanced Computing Center. The DFT calculations were performed using computation resources at the National Energy Research Scientific Computing Center (NERSC). The work at Yale University is partially supported by the National Science Foundation under DMR-2132343.

\appendix

\section{Experimental methods}

Electric resistivity and heat capacity measurements were carried out by using a commercial PPMS (Quantum Design). The electric resistivity was measured by the four-probe method with the current applied in the ac-plane of a \TNS\ single crystal. The specific heat measurement was performed in the temperature range from 200K to 400K where the background signal was recorded in the same temperature range.

To characterize the structural dynamics of \TNS\ with a changing temperature, we carried out XRD measurement with the X-ray energy of 44 keV at the beamline QM2 of the Cornell High Energy Synchrotron Source (CHESS). Needle-like samples were chosen with a typical lateral dimension of $\sim$100 microns, which were then mounted with GE Varnish on a rotating pin. A Pilatus 6M 2D detector is used to collect the diffraction pattern with the sample rotated 360$^{\circ}$ around three different axes at $0.1^\circ $ step and 0.1s/frame data rate. The full 3D intensity cube is stacked and indexed with the beamline software package created by Jacob P. C. Ruff.

Synchrotron-based ARPES measurements were performed at beamline BL5-2 of Stanford Synchrotron Radiation Laboratory (SSRL), SLAC, USA, and BL 7.0.2 (MAESTRO) of Advanced Light Source (ALS), USA. The samples were cleaved \textit{in situ} and measured under the ultra-high vacuum below 3$\mathrm{\times}$10$^{-11}$ Torr. Data was collected by R4000 and DA30L analyzers. The total energy and angle resolutions were 10\,meV and 0.2$\mathrm{^\circ}$, respectively.

\section{First-Principles DFT and Many-Body Simulations}\label{app:theoryDetails}

The \textit{ab initio} calculations presented in Sec.~\ref{sec:theory} are performed using the Quantum ESPRESSO package\,\cite{giannozzi2009quantum}. The structural relaxation is calculated using the r$^{2}$SCAN\,\cite{furness2020accurate} functional with a semiempirical Grimme's DFT-D2 Van-der-Waals correction\,\cite{grimme2006semiempirical}. A 30$\mathrm{\times}$30$\mathrm{\times}$15 \textbf{k}-mesh was used with a 100 Ry wavefunction energy cut-off. The phonon calculations are performed by frozen phonon method using phonopy code\,\cite{phonopy}. A 3$\mathrm{\times}$2$\mathrm{\times}$1 supercell was used. The dimensionless operator $\Delta_{\boldsymbol{q}}^{(\nu)}$ in Eq.\,(\ref{eq:el-ph}) is directly related to the $\nu$-th phonon mode at $\boldsymbol{q}$ point
\begin{equation}
\Delta_{\boldsymbol{q}}^{(\nu)}=\sqrt{\frac{\hbar}{2\omega_{\boldsymbol{q}\nu}}}\sum_{\kappa}\frac{1}{\sqrt{M_\kappa}} \mathbf{e}^{(\nu)}_{\kappa}(\boldsymbol{q}) \cdot \frac{\partial}{\partial \boldsymbol{\tau}_{\kappa}}
\end{equation}
where $\kappa$ labels atoms in the primitive unit cell, $M_\kappa$ is the mass and $\boldsymbol{\tau}_{\kappa}$ is the coordinate of the $\kappa$-th atom, $\mathbf{e}^{(\nu)}_{\kappa}(\boldsymbol{q})$ is $\nu$-th eigenvector of the dynamical matrix at $\boldsymbol{q}$ point on $\kappa$-th atom and $\omega$ is the phonon frequency.

The exact diagonalization (ED) simulations presented in Sec.~\ref{sec:theory} are conducted on an 8-site chain (effectively 16 sites due to the two bands). The diagonalization of the Hamiltonian leads to the ground-state wavefunction $|G\rangle $ and all excited states (denoted as $|m\rangle $, with $|m=0\rangle =|G\rangle $). The finite-temperature spectral function, for each band, is calculated through a canonical ensemble average
\begin{eqnarray*}
A^c\!\left(k,\omega \right)=-\frac{1}{\pi N}\!\sum_{m,\sigma} \frac{e^{-\frac{E_m}{k_B\!T}}}{Z}\mathrm{Im} \langle m|c^\dagger_{k\sigma}\frac{1}{\omega\! +\!\mathcal{H}\!-\!E_m\!+\!i\delta}c_{k\sigma}|m\rangle\\
A^f\!\left(k,\omega \right)=-\frac{1}{\pi N}\!\sum_{m,\sigma} \frac{e^{-\frac{E_m}{k_B\!T}}}{Z}\mathrm{Im} \langle m|f^\dagger_{k\sigma}\frac{1}{\omega\! +\!\mathcal{H}\!-\!E_m\!+\!i\delta }f_{k\sigma}|m\rangle
\end{eqnarray*}
where $Z$ is the partition function. The ensemble-averaged equal-time observables, e.g., the average phonon occupation, are defined in a similar manner
\begin{eqnarray}
\langle O\rangle =\ \frac{1}{Z}\sum_m{e^{-\frac{E_m}{k_BT}}\langle m|O|m\rangle \ \ \ }
\end{eqnarray}
To capture the high momentum resolution comparable to experiments, we employ the twisted average boundary condition (TABC) in the simulation, with 50 equal-spacing phases for the spectral simulation and 30 phases for the equal-time observables. In addition to achieving the momentum resolution, the TABC is known to reduce the finite-size effects, particularly for the model with $q=0$ interactions\,\cite{poilblanc1991twisted}. The ED simulation provides the solution of the full many-body state in 1D and preserves reflection symmetry, i.e. $\langle x\rangle \equiv 0$; however, the fluctuation of displacements $\langle x^2\rangle \neq 0$, reflecting the phonon squeezed state.

\section{Estimation of charge doping}\label{app:doping}

\begin{figure}[!t]
\centering
\includegraphics[width=8.5cm,keepaspectratio=true]{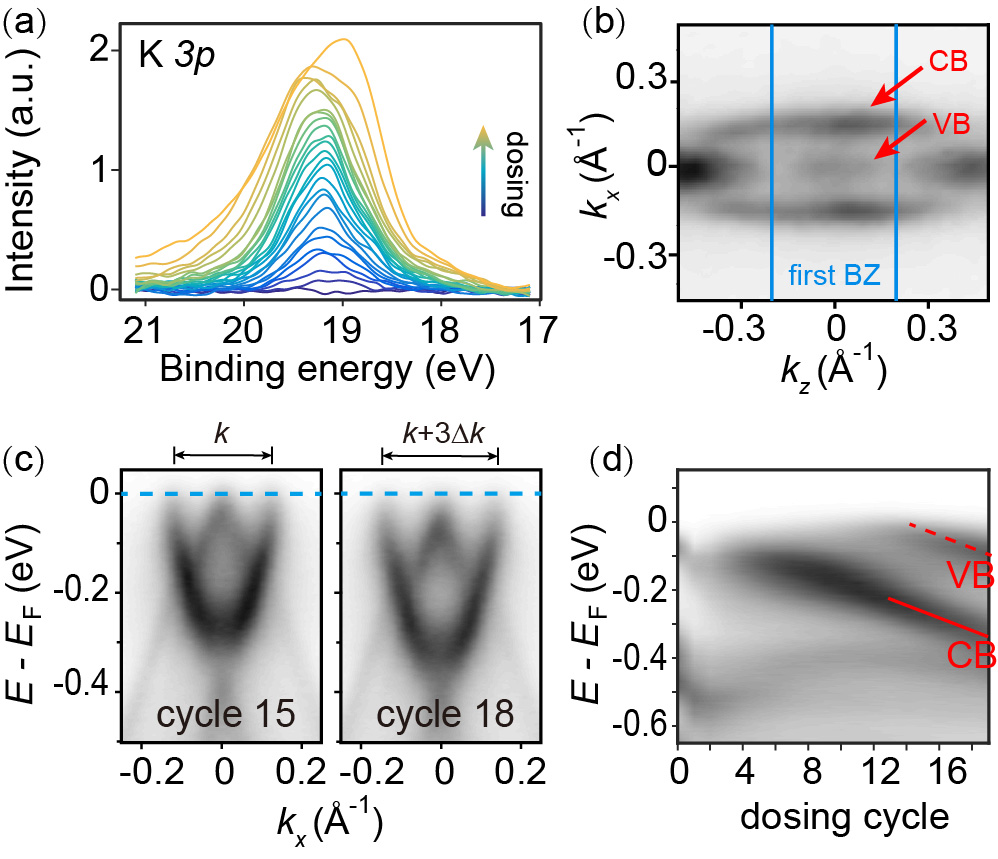}\vspace{-3mm}
\caption{Estimation of electron doping and band shift. (a) Evolution of K $3p$ corelevel peaks with potassium dosing. (b) Constant energy contour at Fermi level after potassium dosing, illustrating the quasi-1D nature of the conduction band (CB) and the valence band (VB).  (c) The change of the momentum width along $X-\mathit{\Gamma}-X$ direction. (d) Evolution of bands at $\mathit{\Gamma}$ point with dosing cycles. 
}
\label{fig:luttinger}
\end{figure} 

The charge doping per unit cell in \TNS\ is quantitatively estimated through the changes of both its band structure and the potassium corelevel peak. As shown in Fig.~\ref{fig:luttinger}(a), the relative amount of potassium deposited on the sample surface is estimated through the increment of the peak area of potassium $3p$ corelevel (Gaussian peak fitting). We then estimate the charge per unit cell using Luttinger theorem, by comparing the area of Fermi surface volume to the size of the Brillouin Zone. \TNS\ has a quasi-1D shape of Fermi surface [Fig.~\ref{fig:luttinger}(b)], therefore only the increment along $X-\mathit{\Gamma}-X$ direction is needed. The change of the momentum width along $X-\mathit{\Gamma}-X$ direction is estimated to be $\mathrm{\sim}$0.01 {\AA}$^{-1}$ for each dosing cycle [Fig.~\ref{fig:luttinger}(c)], which contributes to a Brillouin Zone portion of roughly 0.01 {\AA}$^{-1}$/1.79 {\AA}$^{-1}$=0.56\%. Given the existence of two conduction bands and one valence band, as well as the spin degeneracy, the total amount of dosage for each cycle is estimated to be roughly 0.56\%*6=0.034 electrons per unit cell. We explicitly used the heavy dosing spectra (cycle 14-19), where the conduction and valence bands (CB and VB) move linearly toward higher binding energy. In this region, the influence of interaction or fluctuation is considered to be minimum.

\begin{figure}[!t]
\centering
\includegraphics[width=8.5cm]{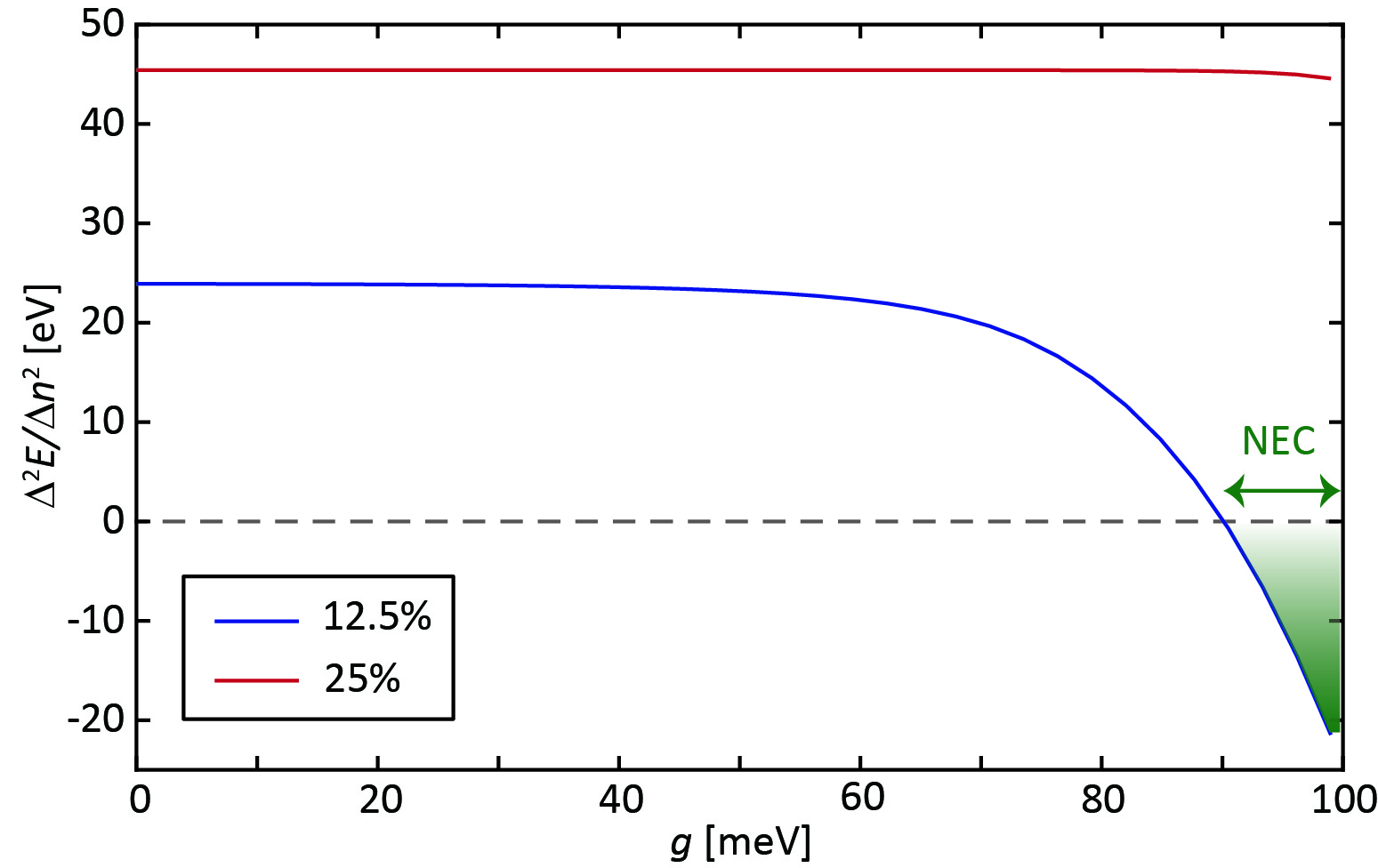}\vspace{-3mm}
\caption{Estimation of electronic compressibility for 12.5\% (blue) and 25\% (red) hole doping, simulated using ED for various EPC strengths. }
\label{fig:NEC}
\end{figure}

\section{Negative Electronic Compressibility}\label{app:NEC}

We estimate the electronic compressibility $\kappa \propto {dn }/{d\mu}$ through the second derivative of total energy with respect to carrier concentration:
\begin{eqnarray}
\kappa^{-1}\propto \frac{d^2E}{dn^2}\approx \frac{E\left(n+ \Delta n\right)+E\left(n-\Delta n\right)-2E\left(n\right)}{\Delta n^2}
\end{eqnarray}
Due to the finite system size in our simulation, we employ a coarse-grained $\Delta n=0.125$. With such a large spacing between neighboring doping levels, the estimated value can be regarded as an average of $\kappa^{-1}$ within a doping interval. The evaluated $\kappa$ turns below zero for strong EPC for 12.5\% doping, while this effect is screened by carriers at larger dopings. Although one cannot compare the quantitative values due to the coarse-grained derivatives, this doping dependence trend matches the low-temperature experiments in Sec.~\ref{sec:lowTemp}. Here, the simulation comes from the $p$-type doping, since the effective mass of the electron pocket is smaller and, accordingly, the NEC doping window is narrower. Our discrete doping regime averages over a 12.5\% doping regime, which smears out the NEC on the electron-doped side.

\section{Hybridization Gap}\label{app:hybridizationGap}

We analyze the hybridization gap of the model Hamiltonian in Eq.~\eqref{eq:manyBodyHam}, to distinguish from the bare band gap $E_g$ caused by the overall band separation. Specifically, we define the hybridization gap $\Delta E_{\rm hyb}$ as twice the off-diagonal element of the mean-field Hamiltonian at $k_{\rm F}$, while the Fermi momentum $k_{\rm F}$ is determined by minimizing $|E_g|$.
If the bare bands overlap, the mean-field Hamiltonian has identical diagonal terms at $k_{F}$, leading to an actual band gap identical to the hybridization gap $\Delta E_{\rm hyb}$; otherwise, $\Delta E_{\rm hyb}$ reflects the contribution of hybridization and the single-particle gap becomes $\sqrt{E_g^2 + \Delta E_{\rm hyb}^2}$.

\begin{figure}[!t]
\centering
\includegraphics[width=8.5cm]{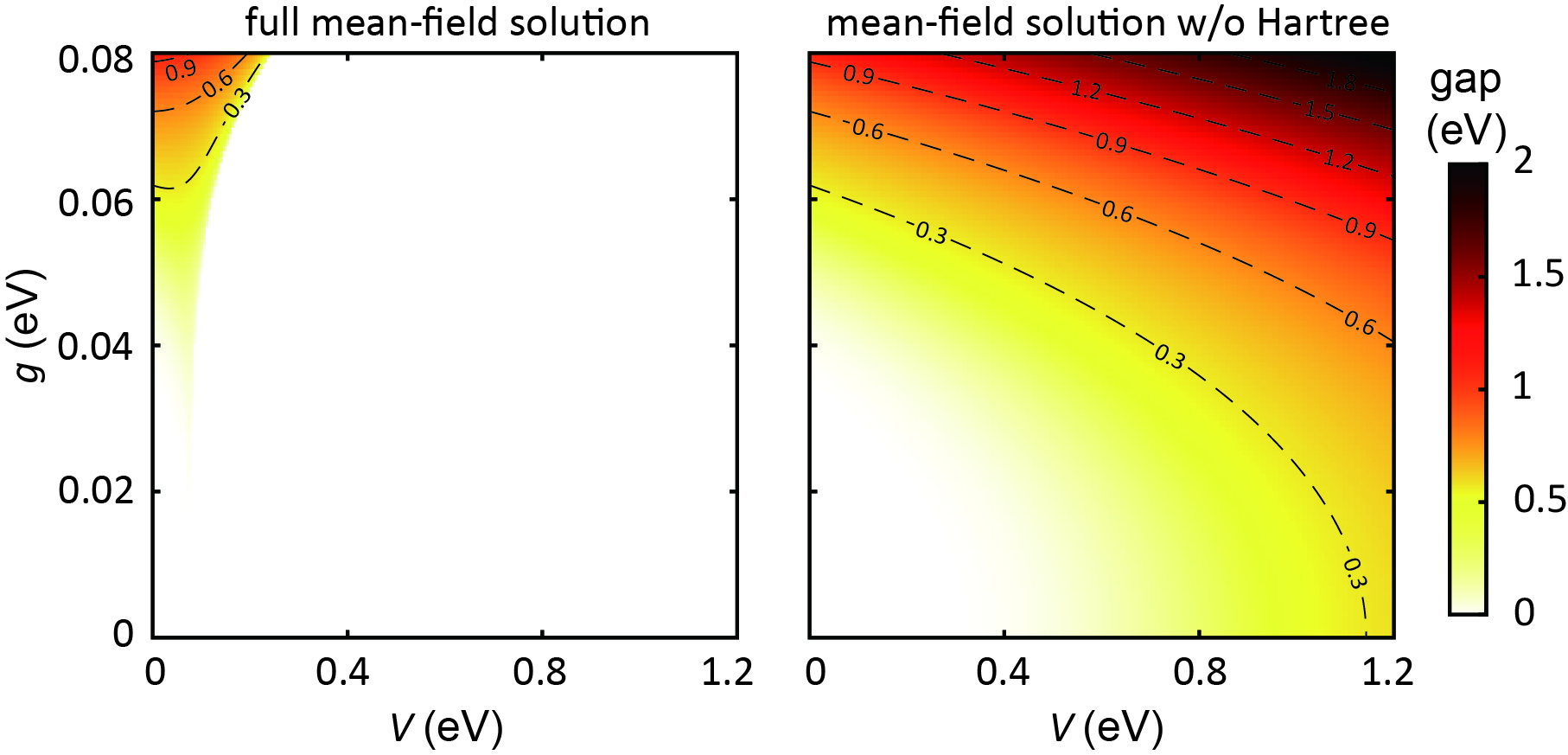}\vspace{-3mm}
\caption{Left panel: The hybridization gap $\Delta E_{\rm hyb}$ induced by the combined electron-electron interaction $V$ and electron-phonon coupling $g$. Right panel: Same as the left panel but artificially removing the Hartree term in the mean-field solution. 
}\label{fig:hybrid}
\end{figure}

The left panel of Fig.~\ref{fig:hybrid} shows the hybridization gap $\Delta E_{\rm hyb}$ determined by the mean-field theory.  A strong Coulomb interaction cannot stabilize a sizeable hybridization gap comparable to experimental results ($\sim 0.3$\,eV), indicating that the critical interaction $V_c$ (defined in Fig.~\ref{fig:Fig9} of the main text) is infinite. In contrast, electron-phonon coupling can efficiently open a sizeable $\Delta E_{\rm hyb}$. This distinction originates from the Hartree term of the Coulomb interaction [same as Eqs.~\eqref{eq:bandShiftc} and \eqref{eq:bandShiftf}], which shifts the two bands away from each other and competes with the exciton formation. To reveal its impact, we further conduct an artificial mean-field simulation by removing the Hartree terms.  This treatment removes the impact of band shifting and maximizes the capability of the Coulomb interaction $V$ to open the hybridization gap. Thus, with no or minimal contribution of the electron-phonon coupling $g$, the Coulomb interaction can open a gap comparable to experiment, but it requires an extremely strong interaction $V\sim 1$\,eV. We set $t_{\rm TaNi} = 0$ for all mean-field simulations except the results in Fig.~\ref{fig:Fig9}(f) of the main text.


\begin{thebibliography}{60}
\bibitem [{pho()}]{phononDiff}%
  \BibitemOpen
  \href@noop {} {}\bibinfo {note} {Note that the lattice order parameter
  $X_{\rm MFT}$ is evaluated through minimizing the mean-field energy rather
  than using the fluctuation-informed displacement $X_{\rm ED}$ in
  Eq.~\eqref{eq:frozenPhonon}.}\BibitemShut {Stop}%
\bibitem [{hyb()}]{hybGap}%
  \BibitemOpen
  \href@noop {} {}\bibinfo {note} {The hybridization gap is not directly the
  order parameter. A more accurate definition is presented in
  Appendix~\ref{app:hybridizationGap}.}\BibitemShut {Stop}%
\end{thebibliography}
%

\pagebreak

\setcounter{figure}{0}
\renewcommand{\figurename}{Fig. S}

\makeatletter
\def\fnum@figure{\figurename\thefigure}
\makeatother

\begin{figure*}[!t]
\centering
\includegraphics[width=5.97in]{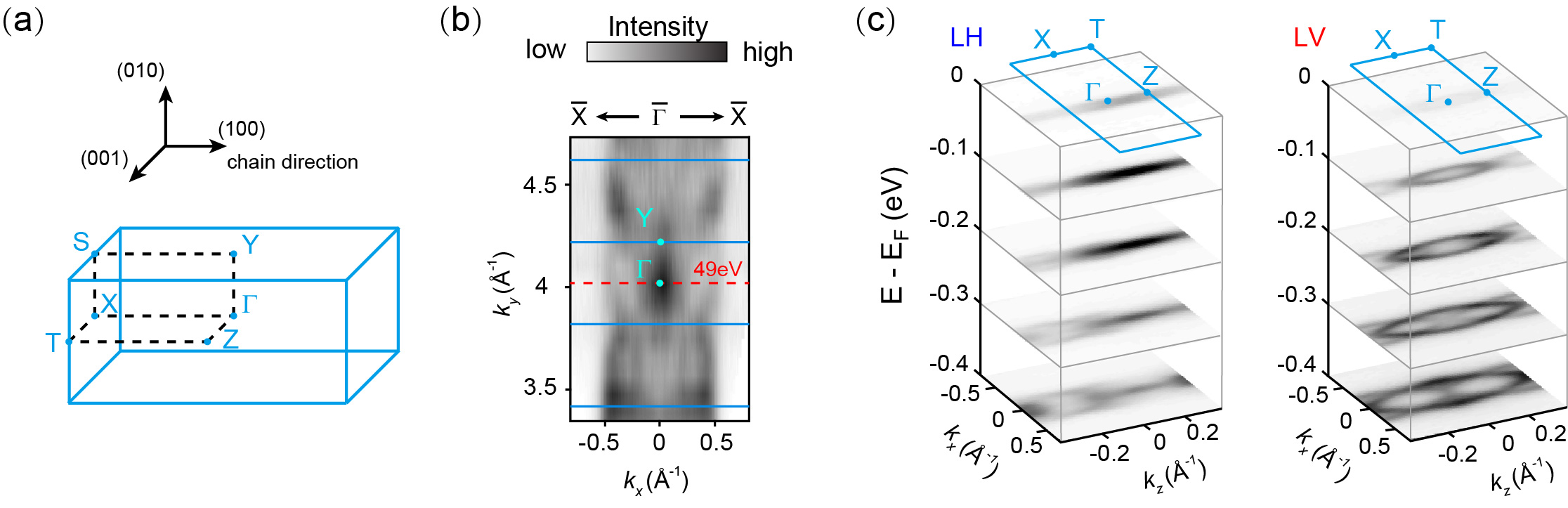}\vspace{-3mm}
\caption{Photoemission data in 3D Brillouin Zone (BZ). (a) Illustration of BZ of \TNS at high-temperature orthorhombic phase. (b) Out of plane photoemission map along the \textit{k${}_{y}$} direction by changing the photon energy of the incident beam. (c) In-plane photoemission map at $\mathit{\Gamma}$ plane (49eV) with linear horizontal (LH) and linear vertical (LV) polarizations of the incident beam. The quasi-1D electronic structure of the system is clearly demonstrated. As the shearing of the Ta atoms in the low-temperature monoclinic phase is very small (less than 1\%), the BZ can be treated as the same for the whole temperature range. 
}
\end{figure*}

\begin{figure*}[!t]
\centering
\includegraphics[width=4.61in]{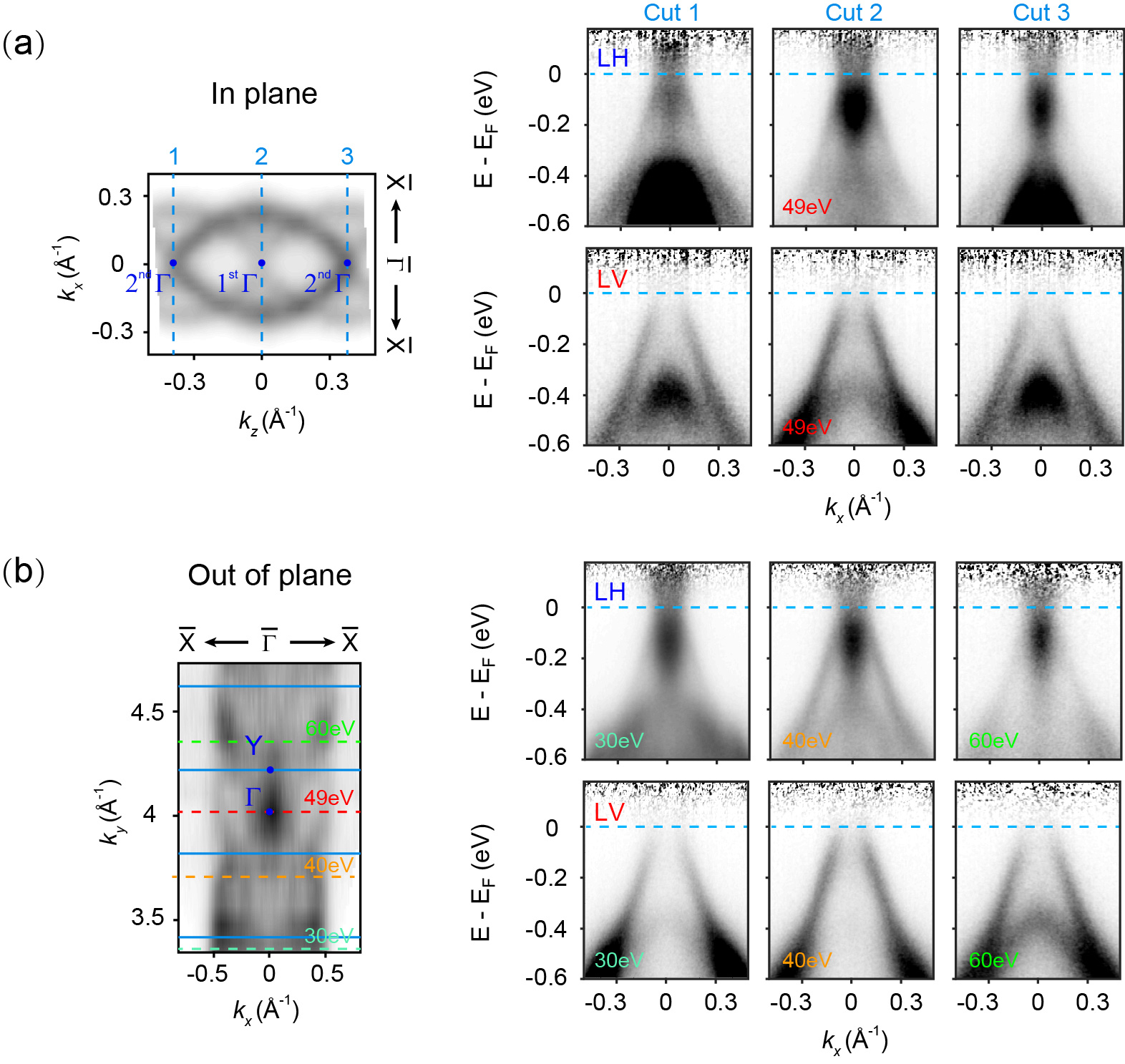}\vspace{-3mm}
\caption{Determination of the pseudogap state at high-temperature phase of \TNS. (a) High symmetry cut $X-\mathit{\Gamma}-X$ at different Brillouin Zones (b) High symmetry cut $X-\mathit{\Gamma}-X$ at different photon energies of the incident beam (different \textit{k${}_{y}$} values). The spectrum weight depletion around $E_F$ is constantly observed. 
}
\end{figure*}

\begin{figure*}[!t]
\centering
\includegraphics[width=6.01in]{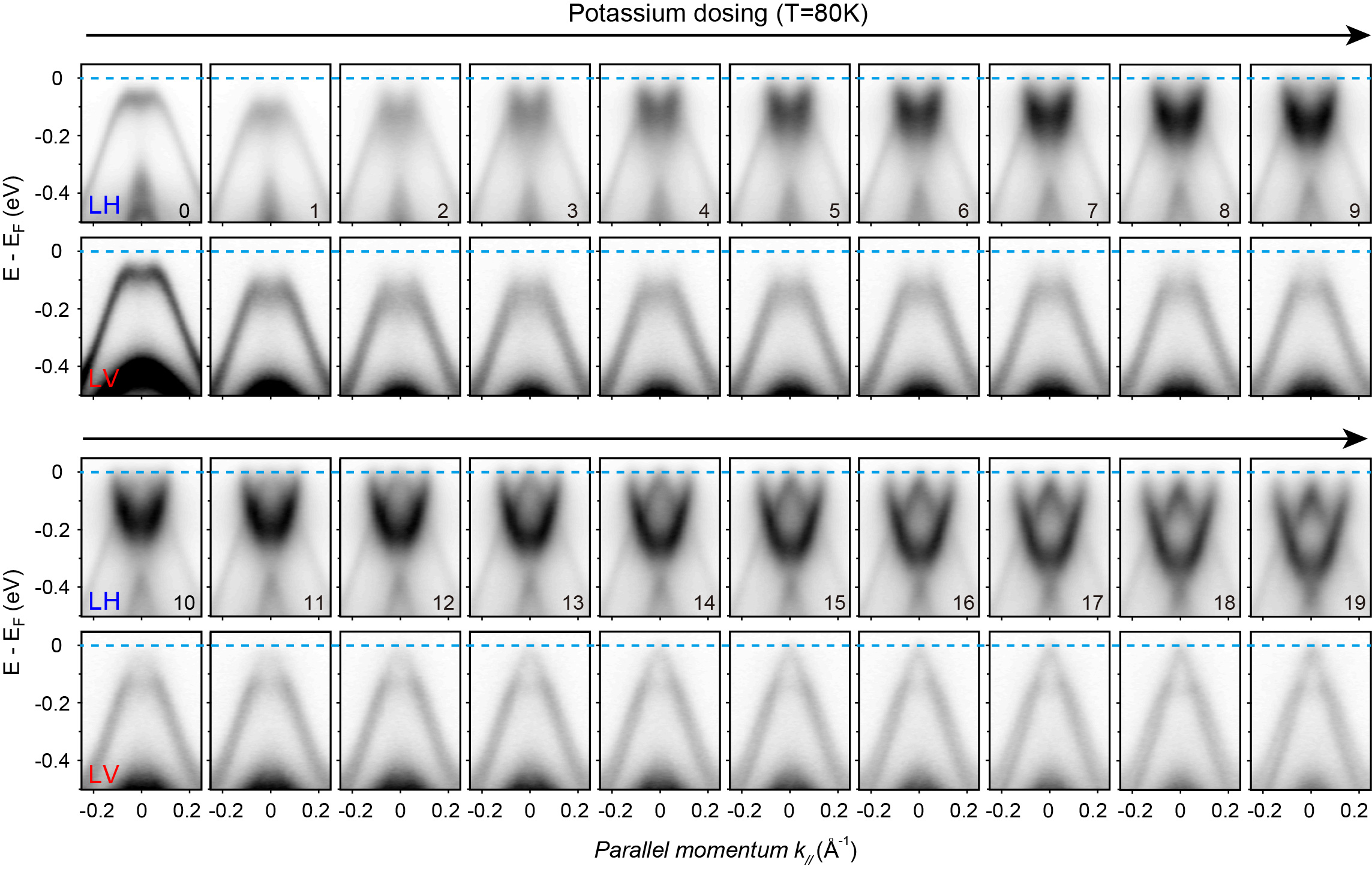}\vspace{-3mm}
\caption{Complete potassium dosing data for both LH and LV polarizations. The dosing experiment was performed with a photon energy of the incident beam at 84eV (also at the $\mathit{\Gamma}$ plane in the 3D Brillouin Zone as illustrated in Fig. S1), due to the availability of both the polarizations of the light at the beamline. Besides the 3-stage evolution illustrated in LH channel (Sec. IV of main text), the LV channel illustrates the evolution of the valence band with the signal mainly from Ni \textit{3d} orbitals. The tip-like valence band top was gradually recovered, consistent with the DFT calculation for the semi-metallic normal state of the system (Sec. III of main text). A second set of bands can be seen in the background (most distinguishable in 16-19), suggesting the phase separation nature of the system during the transition from the monoclinic phase to the orthorhombic phase. Such a behaviour further indicates the structural instability of the system, as a natural result of the presence of fluctuating lattice. 
}
\end{figure*}

\begin{figure*}[!t]
\centering
\includegraphics[width=4.02in, height=3.68in]{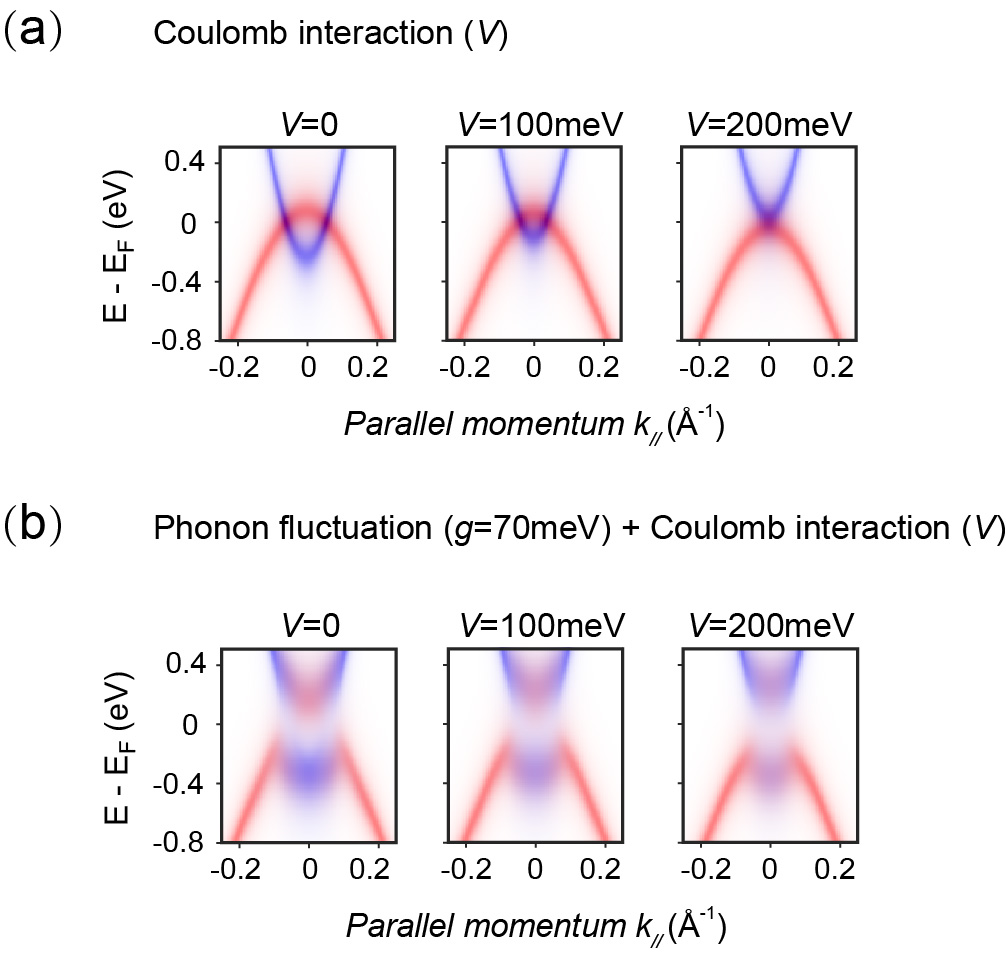}\vspace{-3mm}
\caption{Role of electron-phonon coupling and the direct Coulomb interaction in the system. (a) At the orthorhombic phase without any phonon fluctuations, the increasing Coulomb interaction (increasing \textit{V}) only leads to a relative band shift without mediating exciton formation. (b) With finite \textit{e-ph} coupling ($g=70meV$) and phonon fluctuations, the Coulomb interaction enhances gap opening and spectral depletion. 
}
\end{figure*}

\end{document}